\documentclass[aps,pra,reprint,amsmath,amssymb,nofootinbib]{revtex4-2}
\usepackage{graphicx}
\usepackage[colorlinks=true,allcolors=blue]{hyperref}

\begin{document}

\title{Bell violation with path-entangled number states under realistic detection}

\author{Christoph F. Wildfeuer}
\email{christoph.wildfeuer@fhnw.ch}
\affiliation{FHNW University of Applied Sciences and Arts Northwestern Switzerland, School of Engineering and Environment
CH-5210 Windisch, Switzerland}

\date{\today}

\begin{abstract}
A single photon delocalised over two modes is entangled, and whether that
entanglement can violate a Bell inequality has been disputed for three
decades. In principle it is settled: the surviving objection was the absence
of a shared phase reference, and a local oscillator supplies one, acting as a
shared frame rather than as a measurement setting. In practice it is open, and what remains is a question about the apparatus. Loss does not spoil a parity measurement so
much as rescale it, carrying it into the same one-parameter family of detector
operators that already contains on--off detection. Within that family we give
closed correlators for path-entangled number states, or N00N states, at arbitrary efficiency,
with dark counts, mode mismatch, phase noise and upstream loss. Because loss
and displacement commute up to a rescaling of the displacement amplitude, loss
before and after the displacement is interchangeable, and a symmetric
experiment is governed by one overall efficiency: the probability that a
heralded photon is detected. For a single photon the
Clauser--Horne--Shimony--Holt threshold on that efficiency is $0.8258$ with
on--off detection and $0.9427$ with parity, so the scheme that needs no
photon-number resolution is the more robust, by twelve percentage points of
loss and a factor of five in tolerable dark counts. Efficiency is no longer the obstacle; the mode overlap is. The best demonstrated telecom coupling, a buildable interferometer and today's detectors give an overall efficiency of $0.862$, at which the required overlap is $0.92$. We show that this overlap carries the heralded photon's single-mode weight as well as the local oscillator's mode match, so buying more of it costs heralding efficiency --- and no source has reported the two together.
\end{abstract}

\maketitle

\section{Introduction}
\label{sec:intro}

Send a single photon onto a balanced beam splitter and the state that emerges,
\begin{equation}
  |\Psi\rangle=\frac{1}{\sqrt2}\big(|1\rangle_a|0\rangle_b
                                   -|0\rangle_a|1\rangle_b\big),
  \label{eq:psi1}
\end{equation}
is a superposition of the photon being in one arm and in the other. It is the
$N=1$ member of the family
\begin{equation}
  |\Psi_N\rangle=\frac{1}{\sqrt2}\big(|N\rangle_a|0\rangle_b
                                     -|0\rangle_a|N\rangle_b\big)
  \label{eq:noon}
\end{equation}
of maximally path-entangled number states, or N00N states \cite{Lee02}, used
in quantum imaging, metrology and sensing
\cite{Sanders89,Boto00,Lee02,Kapale05}. Its nonlocality has also been argued
about longer than that of any other entangled state. Tan, Walls and Collett
proposed a Bell test for it in 1991 \cite{TWC91}, and Santos objected the
following year \cite{Santos92}. The argument then ran through Hardy
\cite{Hardy94}, Peres \cite{Peres95}, Vaidman \cite{Vaidman95} and
Greenberger, Horne and Zeilinger \cite{GHZ95} in the mid-nineties. It was
still being conducted in 2005 and 2006, between van Enk and Drezet
\cite{vanEnk05,Drezet06,vanEnk06}. It is not unusual, even now, to hear that a
single photon with vacuum in the other arm cannot violate a Bell inequality at
all.

Three questions are worth separating; running them together has kept the
argument going. Is the state entangled? Is it Bell-nonlocal in principle? And
can a given experiment reach that nonlocality? Only the third is open.

The first has an answer: the state of Eq.~(\ref{eq:psi1}) is entangled. Van
Enk's argument settles it in three lines \cite{vanEnk05}. Put an atom in a
cavity in each arm; local absorption maps Eq.~(\ref{eq:psi1}) onto
$(|g\rangle_a|e\rangle_b+|e\rangle_a|g\rangle_b)/\sqrt2$, with $g$ and $e$ the
atomic ground and excited states, and local operations cannot create
entanglement, so the photonic state had it already. This settles the first
question and not the third: the local superselection rule of the next
paragraph applies just as well to atomic excitation number, so that
superposition is as inaccessible in isolation as the optical one.

The second also has an answer. A Bell violation is not an operational measure
of entanglement. Entanglement, and Bell nonlocality form a strict
hierarchy \cite{Werner89,WJD07}: there exist entangled mixed states admitting
a local hidden-variable model for projective measurements \cite{Werner89}, and
indeed for general measurements \cite{Barrett02}, so violation is strictly the
stronger property. For \emph{pure} states, however, the hierarchy collapses.
By Gisin's theorem every pure entangled state of two qubits violates the CHSH
inequality for suitable observables \cite{Gisin91}, a result later extended to
arbitrary bipartite dimension \cite{GisinPeres92} and to many parties
\cite{PopescuRohrlich92}. The two-qubit case suffices here:
Eq.~(\ref{eq:psi1}) has Schmidt rank two and is supported entirely in
$\mathrm{span}\{|0\rangle,|1\rangle\}^{\otimes2}$, so any $\pm1$-valued qubit
observable on that subspace extends to a $\pm1$-valued observable on the full
Fock space without changing any expectation value. Its Bell nonlocality is
therefore not in question either --- provided arbitrary local observables are
permitted.

The third is in question, and here the sceptics had a real point, later made
precise by Bartlett, Rudolph and Spekkens \cite{BRS07}. Gisin's observables
are arbitrary on each mode, and arbitrary means coherences between different
photon numbers --- exactly what a local particle-number superselection rule
forbids without a reference. Under such a rule the state cannot violate a Bell
inequality --- but only when the two parties do not share a phase reference.
Supply one and the restriction is not so much lifted as paid for: the joint
operation on signal and reference still conserves photon number, and induces
on the signal alone exactly the operation forbidden in isolation. In the
optical setting the phase reference is a local oscillator, and the measurement
it enables is a displacement in phase space. This is what the
Tan--Walls--Collett scheme was reaching for, and what Hardy's modification of
it \cite{Hardy94} and every subsequent proposal implements.

So the conceptual question is closed; an experimental one remains. Every measured violation with this state so far has needed
post-selection or a fair-sampling assumption
\cite{Hessmo04,Babichev04,DAngelo06}; the post-selection-free experiments have
demonstrated entanglement \cite{Monteiro15} and Einstein--Podolsky--Rosen
steering \cite{Fuwa15,Guerreiro16}, not a Bell violation. A Bell violation
trusts neither device, steering trusts one, entanglement witnessing trusts
both, so the efficiency each demands falls down that list. The gaps are large.
Steering tolerates arbitrarily high loss given enough settings
\cite{Bennet12}; no Clauser--Horne--Shimony--Holt test can go below the
Eberhard limit \cite{Eberhard93}. Guerreiro \textit{et al.}\ reached an
overall efficiency of about $0.53$ where $0.84$ is needed at their own
reported mode overlap \cite{Guerreiro16}; a stricter reading of their budget
gives $0.42$ (Sec.~\ref{sec:budget}). A recent two-copy experiment, in which
one photon supplies the phase reference for the other, reports $|S|=2.71$ when
post-selected on distributed detections and $|S|=2.23$ without post-selection
\cite{Kun25}; the second is the figure comparable to what follows. No overall
Bell-test efficiency is quoted there and no loophole-free claim is made. No
loophole-free Bell test with single-photon path entanglement has been
reported.

This paper asks what the numbers actually are. It extends the analysis of
Ref.~\cite{Wildfeuer07}, which treated both detection schemes for these states
assuming ideal detectors, to detectors that are not ideal.

We work with the two schemes used there: simple on--off detection, which
measures the $Q$ function \cite{BW99}, and correlated parity detection, which
measures the Wigner function \cite{BW99,Wildfeuer07}. We recall first,
following Refs.~\cite{BW96,WV96,BRWK99}, that under loss they are not two
schemes but one operator family, at two efficiencies, and show that they are
nevertheless inequivalent as Bell observables even where the operators
coincide. We give the correlator for Eq.~(\ref{eq:noon}) at arbitrary $N$ and
arbitrary efficiency, and then compute the Clauser--Horne--Shimony--Holt
maximum for Eq.~(\ref{eq:psi1}) as a function of detection efficiency,
dark-count rate, local-oscillator mode overlap, local-oscillator phase noise,
and loss upstream of the measurement. We show that detector efficiency and
upstream loss enter only through their product, for any $N$ and either
detector, so that the whole budget of a symmetric experiment collapses onto
one overall efficiency. We also show that the mode overlap in that budget is
not a property of the local oscillator alone: it carries the heralded photon's
single-mode weight, and therefore trades against the very heralding efficiency
the budget needs. The result is a budget an experimentalist can check a lab
against.

\section{Is a single photon nonlocal?}
\label{sec:debate}

\subsection{The claim and the objection}

Tan, Walls and Collett \cite{TWC91} mixed each output of the beam splitter
with a weak coherent field at a separate station and formed a correlation
function from the intensities, reporting a Clauser--Horne--Shimony--Holt
violation. Santos objected that the correlations admitted a local
hidden-variable model \cite{Santos92}, and that exchange has since been
settled against the original claim: Das \textit{et al.}\ exhibited such a
model throughout the regime in which the violation was claimed
\cite{Das21,DasNoGo}. That scheme is therefore not a valid demonstration and
we do not build on it. The objection was to the correlation function, not to
the state. Hardy \cite{Hardy94} made the local-oscillator amplitude itself a
measurement setting, so that the two settings at each station differ by the
auxiliary field being on or off. The same authors who dismantled
Tan--Walls--Collett judge Hardy's scheme valid, identifying the varying
local-oscillator strength as what enables the violation \cite{Das21}, and
later call the resulting non-classicality unquestionable
\cite{Schlichtholz26}. Hardy's proposal is the ancestor of the displacement
measurements used throughout this paper. The objections that followed were not
all the same objection. Peres \cite{Peres95} and Vaidman \cite{Vaidman95}
accepted that a violation occurs and disputed only the label --- Peres
attributing it to a two-particle component created by the detection process,
Vaidman to the additional systems at the two stations and, on pre- and
post-selection grounds, denying there is a paradox at all. Neither questions
the correlations. Greenberger, Horne and Zeilinger \cite{GHZ95} also called
the effect a multiparticle one in disguise, but their central objection was to
the physics: Hardy's coherent superpositions of vacuum and one photon are
unobservable in isolation, and can enter a real experiment only if other
photons are present to keep track of the phase. That is the substantive
objection, and in modern language it is a superselection rule.

\subsection{Superselection rules, and what a reference buys}

A \emph{local} particle-number superselection rule --- the global rule by
itself forbids nothing here --- prevents each party from preparing or
measuring coherences between different photon numbers. That local restriction
is not an independent law of nature; it is operationally the statement that no
shared phase reference is available \cite{BRS07}. Under it the entanglement of
Eq.~(\ref{eq:psi1}) survives as a fact about its preparation: the state still
cannot be made by unrestricted local operations and classical communication.
What is gone is access to it. In Wiseman and Vaccaro's language \cite{WV03}
the entanglement of modes is nonzero while that of particles is zero, and a
reference \emph{activates} the former --- an analogy with bound entanglement
made precise by Bartlett \textit{et al.}\ \cite{BDSW06}.

The resolution is older than the objection: whether a superposition of a
superselected quantity is preparable depends on the availability of a suitable
reference \cite{AS67}. Bartlett, Rudolph and Spekkens state the consequence
for exactly our state \cite{BRS07}: it cannot violate a Bell inequality
\emph{when the two parties do not share a phase reference}, and combined with
a shared reference it gives something that can. The superselection rule is not
a law forbidding the experiment; it is a bookkeeping device that tracks
whether the phase reference has been counted. The experiment needs only a
\emph{relative} phase between two beams drawn from one source, not the
disputed \emph{absolute} optical phase \cite{RudolphSanders01,vanEnkFuchs02}.

Dunningham and Vedral supplied the operational version \cite{DV07}: they
reproduced Hardy's argument using only number states and phase-averaged
mixtures of coherent states, with no superposition of vacuum and one photon
anywhere in the description, taking the local oscillators from the reference
state. They conclude that the Greenberger--Horne--Zeilinger objections are
met, and that nonlocality is better seen as a property of quantum fields than
of particles. Ashhab \textit{et al.}\ reached a weaker version with an
ancillary Bose--Einstein condensate as the reference \cite{Ashhab09}, a
\emph{probabilistic} violation in which a finite fraction of runs violate
although the average does not. One limit on this line of work should be
recorded. The violation Cooper and Dunningham \cite{CD08} claimed for
completely independent reference states rests on a calculational error and
does not occur \cite{DasComment}. The resolution stands for schemes where each
party holds a genuine local oscillator, the case treated here and realised in
Refs.~\cite{Monteiro15,Guerreiro16,Caspar22}; we do not rely on the
independent-reference variant.

\subsection{What the displacement measurement supplies}

Two properties of the measurement answer the objections above.

First, the local oscillator is the phase reference whose absence was the
substance of the superselection objection. Mixing the signal mode with a
strong coherent field on a highly transmissive beam splitter implements a
displacement $\hat D(\alpha)$ \cite{Paris96,BW96,WV96}, precisely the
operation a particle-number superselection rule forbids without a reference.
This closes the gap Greenberger, Horne and Zeilinger pointed at, and the
closure is what Bartlett, Rudolph and Spekkens make formal.

An ideal shared frame lifts the restriction outright and returns the full
observable algebra that Gisin's theorem draws on. What a \emph{local
oscillator} returns is less: a coherent field and photodetection realise the
displaced number-diagonal family and nothing more.
Section~\ref{sec:thresholds} shows that this family already violates the
Clauser--Horne--Shimony--Holt inequality, with $\max|S|=2.6884$, and that is
what closes the argument. The local oscillator is treated throughout as a
classical field; at the few hundred photons per pulse the optimum uses, the
corrections from a bounded reference \cite{BRST09} are far below anything else
in the budget.

Second, every trial produces an outcome. An on--off detector either clicks or
does not, and a number-resolving detector returns some $n$ whose parity is
$\pm1$; neither outcome is discarded and no event is conditioned away. There
is no post-selection at the measurement, hence no fair-sampling assumption and
no detection loophole. This property distinguishes the modern experiments
\cite{Monteiro15,Guerreiro16} from the early ones \cite{Babichev04}, whose
authors were explicit that their tests assumed fair sampling.

The trial itself, however, is defined by a herald. This is an
\emph{event-ready} test in Bell's sense, of the kind
Refs.~\cite{Hensen15,Rosenfeld17} realised with matter qubits. The idler is
detected at the source, has no access to either setting, and fixes the trial
list before any setting is applied; conditional on it, both stations always
return $\pm1$. The conditioning is therefore legitimate, and the efficiency
that matters is the conditional one --- the probability that a heralded photon
reaches a detector --- which is exactly the $\mu$ of
Sec.~\ref{sec:invariance}, with the heralding efficiency carried by the
upstream factor. The herald detector's own efficiency does not enter $\mu$
(Sec.~\ref{sec:multiphoton}), though its dark counts do.

Given both, a violation would mean what a violation is supposed to mean,
provided four further things hold. No trial is dropped after the herald ---
not for a latched detector, a phase excursion, or a modulator that has not
settled, and not in a setting-dependent way. The settings are drawn locally at
random and each choice is spacelike separated from the distant outcome. Each
choice is also spacelike separated from the emission, so that it is
independent of the hidden variable. And the statistical analysis does not
assume independent trials. Section~\ref{sec:exp} treats all four, and shows
that a natural layout gets the first and third wrong.

Nothing on that list requires knowing $\eta$, $\xi$ or $p$: those are design
quantities, and a measured $|S|>2$ does not depend on any of them. Nor does
anything require the detection efficiency to be independent of the setting.
The Clauser--Horne--Shimony--Holt bound is derived from three assumptions ---
$\pm1$ outcomes on every trial, settings independent of the hidden variable,
and locality --- and none of them mentions efficiency. A setting-dependent
transmission is part of the local measurement that the setting labels, so a
local model equipped with one is still bound by $|S|\le2$; it biases the
comparison between the measured $S$ and the model, but it cannot manufacture a
violation. The hazard in the same situation is setting-dependent \emph{trial
loss}, which is the first condition. A dropped trial leaves an ensemble that
the setting helped to select, so the premise that every trial returns $\pm1$
fails, and the bound fails with it. The obstacle is not conceptual. It is that
the measured Clauser--Horne--Shimony--Holt value $|S|$, defined in
Eq.~(\ref{eq:chsh}) below, must exceed $2$ with detectors that are not
perfect, and the rest of this paper is about how good they have to be.

A standard bipartite, two-setting Bell test on Eq.~(\ref{eq:psi1}) using only
passive linear optics cannot work: passive optics and photodetection realise
only number-diagonal local measurements, which is exactly what a reference is
needed to escape. Abiuso \textit{et al.}\ state the conclusion
\cite{Abiuso22}. Within the scheme considered here the displacement is an
active operation and it is not optional. Elsewhere it can be traded for extra
modes: the same authors show that single-photon entanglement generates
nonlocal correlations in a network of beam splitters and photodetectors with
no displacement at all.

\subsection{A frame is not a setting}

Sharing a phase reference does not compromise the independence of the
settings, and the distinction is between a frame and a choice. Bell's theorem
requires that the setting choices be statistically independent of the hidden
variable and be made locally. It does not require that the two parties share
nothing, any more than a polarisation test requires that they disagree about
what ``vertical'' means. The local oscillator phase plays the same role here.
It fixes the direction of the real axis in phase space, and within that frame
Alice's free choice is between $\alpha$ and $\alpha'$.

A local hidden-variable model may be granted the \emph{value} of the relative
phase outright: Bell's bound is derived allowing the hidden variable to be
arbitrary shared classical data established in the common past, so one more
such variable cannot raise it. What it cannot be granted is the
local-oscillator field itself. Paterek \textit{et al.}\ make the distinction
precise \cite{Paterek11}: with a particle-number superselection rule in force,
a shared reference frame enables a Bell violation if and only if it was
prepared jointly, and not by local operations and classical communication. The
two oscillators must therefore be phase-locked, drawn from a common source,
which is what the experiments do. Two things should be kept apart here. Joint
preparation is what makes a violation \emph{obtainable}; it is not what makes
a measured violation \emph{meaningful}, since a local model is not itself
bound by a superselection rule and $|S|\le2$ holds for it whatever the
oscillators' provenance. Within the restricted theory the statement does have
content: a violation certifies that the shared reference was not preparable by
local operations and classical communication.

Sharing a reference does require a constraint on the layout, the one every
Bell test faces. The displacement must be applied \emph{locally}, at each
station, after the local oscillator has arrived, with each choice spacelike
separated from the distant outcome. In the experiments of
Refs.~\cite{Monteiro15,Guerreiro16,Caspar22} the displacement amplitude and
phase are set by a polarisation controller at the station, which is what one
wants. What would genuinely invalidate the test is setting the displacement
centrally, at the source, since the choice would then lie in the common past
of both outcomes. The co-propagating local oscillator therefore delivers the
frame, not the setting.

A violation certifies that no local model reproduces the correlations
\emph{given} that a phase reference is available to both parties; on the
accounting of Ref.~\cite{BRS07} the nonlocality belongs to the state together
with the reference. Van Enk's reply applies \cite{vanEnk05,vanEnk06}: the same
is true of polarisation entanglement, where the shared frame is equally
physical and is merely carried by degrees of freedom the apparatus never
detects. The demand on the frame can in any case be made very weak. Brask,
Chaves and Brunner showed that displacement-based tests on this state violate
local realism even when the relative phase fluctuates from run to run,
provided only that its distribution is not uniform \cite{BCB13}; a completely
unknown phase is the no-reference case, and it admits no violation. The
oscillators are still drawn from a common source there; what is given up is
knowledge of the relative phase, not its joint preparation. Their scheme needs
more settings and more efficiency than the aligned one, so the thresholds
below assume a controlled relative phase.

\section{One detector, one parameter}
\label{sec:family}

The measurement is unbalanced homodyning \cite{Paris96,BW96,WV96,BW99}. Each
mode is mixed with a strong coherent state on the displacement coupler, a beam
splitter of transmissivity $T_d\to1$. That acts on the signal as a
displacement $\hat D(\alpha)$, with $\alpha=\gamma\sqrt{1-T_d}$ held finite
and $\gamma$ the local-oscillator amplitude. The displaced mode is then
detected. Two detectors have been used. An on--off detector distinguishes only
vacuum from everything else, giving the $Q$ function \cite{BW99}; a
number-resolving detector records the parity $(-1)^{\hat n}$, giving the
Wigner function \cite{Royer77,MoyaCessa93,BW99,Wildfeuer07}.

Under loss the two are one operator family. A detector of quantum efficiency
$\eta$ is an ideal detector preceded by a beam splitter of transmissivity
$\eta$. The loss channel $\Phi_\eta$ has Kraus operators $\hat
K_k=\sqrt{(1-\eta)^k/k!}\,\eta^{\hat n/2}\hat a^k$, where $\hat a$ is the
annihilation operator and $\hat n=\hat a^\dagger\hat a$ the photon number.
These satisfy $\sum_k\hat K^\dagger_k\hat K_k=\hat 1$. Since $\eta^{\hat
n/2}z^{\hat n}\eta^{\hat n/2}=(\eta z)^{\hat n}$ for any $z$, the adjoint
channel is diagonal in the number basis and the sum over Kraus indices is
binomial,
\begin{equation}
  \Phi^\dagger_\eta\big[z^{\hat n}\big]
  =\sum_{k}\tbinom{\hat n}{k}(1-\eta)^k(\eta z)^{\hat n-k}
  =(1-\eta+\eta z)^{\hat n}.
  \label{eq:lossgeom}
\end{equation}
Both detectors follow from it. Parity is $z=-1$, giving
\begin{equation}
  \Phi^\dagger_\eta\big[(-1)^{\hat n}\big]=(1-2\eta)^{\hat n},
  \label{eq:lossparity}
\end{equation}
and an on--off detector, whose silent outcome is $z=0$ since $0^{\hat
n}=|0\rangle\langle0|$, stays silent with probability $\langle(1-\eta)^{\hat
n}\rangle$. Both are contained in one operator,
\begin{equation}
  \hat M(x,\alpha)=\hat D^\dagger(\alpha)\,x^{\hat n}\,\hat D(\alpha),
  \qquad
  x=\begin{cases} 1-2\eta, & \text{parity},\\[2pt]
                  1-\eta,  & \text{on--off},\end{cases}
  \label{eq:M}
\end{equation}
whose expectation value is, up to normalisation, the $s$-ordered
quasiprobability function with $s=-(1+x)/(1-x)$. Explicitly, $\hat
M(x,\alpha)=\tfrac12(1-s)\,\hat T(-\alpha,s)$ with $\hat T$ the
Cahill--Glauber operator \cite{CG69}; note the argument $-\alpha$, which is
immaterial in Eq.~(\ref{eq:Pigen}) but matters when matching single-mode
conventions.

That detector efficiency and quasiprobability ordering are one variable is due
to Banaszek and W\'odkiewicz \cite{BW96,BRWK99} and to Wallentowitz and Vogel
\cite{WV96}; the underlying $s$-ordered family is Cahill and Glauber's
\cite{CG69}. Lee, Jeong and Jaksch went further \cite{LJJ09}: their Eq.~(8)
identifies detector inefficiency with a shift along the ordering axis, and it
contains both of the branch relations above. They also drew the comparative
conclusion that concerns us here --- that the $Q$ branch tolerates the most
loss --- and quoted about $83\%$ for it. Both branches run past the $Q$
endpoint under loss, but not at the same rate:
$s_{\text{on--off}}(\eta)=s_{\text{parity}}(\eta/2)$, so parity sweeps the
whole of $s\le0$ while on--off sweeps only $s\le-1$ and moves twice as fast in
$\eta$. The two detector types do not interpolate between the Wigner and $Q$
endpoints; they traverse the same axis differently.

Setting $\eta=1/2$ in the parity case gives $x=0$, where $x^{\hat n}$
collapses to $|0\rangle\langle0|$: a parity detector of efficiency one half
returns the displaced vacuum projector. This is one substitution into the
parity-branch relation $s=1-1/(\eta T_d)$ \cite{BRWK99} --- the on--off branch
is $s=1-2/(\eta T_d)$ --- or into Eq.~(8) of Ref.~\cite{LJJ09}, and we claim
no priority for it. It fixes the geometry of what follows. What is spent as
the efficiency falls is not the detector's information --- the counting
distribution is still there --- but the usefulness of \emph{parity} as the
function that dichotomises it, and at $\eta=1/2$ there is none of that left.

Both schemes are tested with the same pair of inequalities. Let $A_a,A_b=\pm1$
be the recorded outcomes at settings $\alpha$ and $\beta$ and
$E(\alpha,\beta)=\langle A_aA_b\rangle$ their correlation. The
Clauser--Horne--Shimony--Holt combination \cite{CHSH69} is
\begin{equation}
  S=E(\alpha,\beta)+E(\alpha',\beta)+E(\alpha,\beta')-E(\alpha',\beta'),
  \qquad |S|\le2
  \label{eq:chsh}
\end{equation}
for any local hidden-variable model. The Clauser--Horne form \cite{CH74} uses
probabilities in place of correlations: with $P(\alpha,\beta)$ the joint
probability of the outcome $+1$ at both stations and $P(\alpha)$, $P(\beta)$
the corresponding marginals,
\begin{align}
  \mathrm{CH}={}&P(\alpha,\beta)+P(\alpha',\beta)+P(\alpha,\beta')\nonumber\\
              &-P(\alpha',\beta')-P(\alpha)-P(\beta),
  \label{eq:ch}
\end{align}
with $-1\le\mathrm{CH}\le0$ locally. Equation~(\ref{eq:ch}) is the inequality
of Ref.~\cite{Wildfeuer07} with the primed and unprimed settings of the first
station interchanged.

Return now to the coincidence at $\eta=1/2$. It does \emph{not} follow that
the two schemes are the same experiment: the operator $\hat M$ coincides at
$x=0$, but the $\pm1$-valued Bell observables do not. A parity detector
records $A=(-1)^{n}$ with $\langle A\rangle=\langle\hat M\rangle$. An on--off
detector assigns $+1$ to no click, whose POVM element is $\hat M$, so that
$\langle A\rangle=2\langle\hat M\rangle-1$. The CHSH combination is not
invariant under that affine relabelling, because the local bound of $2$ is
tied to $A=\pm1$. Numerically the difference is large: at $x=0$ a parity
detector gives $\max|S|=0.8881$, attained with all four amplitudes equal to
$1/\sqrt2$, the three added terms at relative phase $\pm135^\circ$ and the
subtracted one at $-45^\circ$, against $2.6884$ for an ideal on--off detector.
The two estimate the same quasiprobability; they are not interchangeable in a
Bell test, and the rest of this paper is about the gap.

\section{Correlators with imperfections}
\label{sec:corr}

\subsection{\texorpdfstring{Arbitrary $N$}{Arbitrary N} and arbitrary efficiency}

Write $\lambda=1-x$. With $|u\rangle$, $|v\rangle$ coherent states, the Fock
matrix elements of $\hat M$ follow from the generating function
\begin{equation}
  e^{(|u|^2+|v|^2)/2}\langle u|\hat M|v\rangle
  =e^{-\lambda|\alpha|^2-\lambda\alpha u^*-\lambda\alpha^*v+xu^*v},
  \label{eq:gen}
\end{equation}
giving $\langle0|\hat M|0\rangle=e^{-\lambda|\alpha|^2}$, $\langle N|\hat
M|0\rangle=(-\lambda\alpha)^Ne^{-\lambda|\alpha|^2}/\sqrt{N!}$ and $\langle
N|\hat M|N\rangle=x^NL_N(-\lambda^2|\alpha|^2/x)e^{-\lambda|\alpha|^2}$, with
$\langle0|\hat M|N\rangle$ the conjugate of the second, since $x$ is real on
both branches and $\hat M$ is Hermitian. Taking the expectation of $\hat
M(x,\alpha)\otimes\hat M(x,\beta)$ in Eq.~(\ref{eq:noon}) and collecting the
diagonal and interference terms, the correlated measurement is
\begin{align}
  \Pi_x(\alpha,\beta)={}&\tfrac12 e^{-\lambda(|\alpha|^2+|\beta|^2)}\nonumber\\
   &\times\Big[x^N\big(L_N(-\tfrac{\lambda^2|\alpha|^2}{x})
                +L_N(-\tfrac{\lambda^2|\beta|^2}{x})\big)\nonumber\\
   &\quad-\tfrac{\lambda^{2N}}{N!}\big((\alpha^*\beta)^N
     +(\alpha\beta^*)^N\big)\Big].
  \label{eq:Pigen}
\end{align}
At $x=-1$, $\lambda=2$ this is the correlated parity expectation
$\Pi(\alpha,\beta)$ of Ref.~\cite{Wildfeuer07}, whose two-mode Wigner function
is $W=4\Pi/\pi^2$. Here $L_N$ is the Laguerre polynomial. Its argument is
singular at $x=0$ but the product is not:
$x^NL_N(-\lambda^2|\alpha|^2/x)=\sum_{k=0}^{N}
\binom{N}{N-k}(\lambda^2|\alpha|^2)^kx^{N-k}/k!$ is a polynomial in $x$,
regular on the whole interval.

\subsection{Dark counts, mode mismatch, phase noise, upstream loss}

\emph{Dark counts.} Let each detector register spurious counts with Poisson
mean $\nu$ per window. Since $\langle(-1)^{n+k}\rangle=e^{-2\nu}
\langle(-1)^{n}\rangle$ for Poissonian $k$, each local parity acquires a
factor $e^{-2\nu}$, and because the Bell combination is bilinear in the local
observables this is an exact rescaling,
\begin{equation}
  S(\nu)=e^{-4\nu}S(0).
  \label{eq:dark}
\end{equation}
Settings optimal without dark counts stay optimal with them. For an on--off
detector a dark count forces a click, so the no-click probability is
multiplied by $e^{-\nu}$, the Poisson probability of no spurious event. That
is not a rescaling of $S$ and is handled numerically. In both cases $\nu$ is a
Poisson mean per detector per window.

\emph{Mode mismatch.} Let the local oscillator have amplitude overlap $\xi$
with the signal mode. The non-overlapping fraction is a coherent state of
amplitude $\sqrt{1-\xi^2}\,\alpha$ in a mode the signal never occupies, on
which the detector reports $e^{-\lambda(1-\xi^2)|\alpha|^2}$, while the signal
mode sees a displacement $\xi\alpha$. The parameter $\xi$ is not a property of
the local oscillator alone. A heralded photon is in general \emph{not} in a
single spectral mode: heralding on a bucket idler detector leaves the signal
in a mixture over the Schmidt modes of the pair, of which the local oscillator
addresses one. The components in the others are displaced by nothing and
interfere with nothing, but the detector, being a bucket over all modes, still
sees them. The correlator is linear in the state, and $\xi$ enters it only as
$\xi^2$ multiplying the $p\lambda^2$ group. At $N=1$ that mixture is therefore
\emph{exactly} equivalent to a pure single-mode photon with
\begin{equation}
  \xi^2=\xi_{\text{LO}}^2\,w,
  \label{eq:xieff}
\end{equation}
$w$ being the weight of the heralded photon in the mode the local oscillator
addresses and $\xi_{\text{LO}}$ that oscillator's own mode match. More
generally, if the signal occupies Schmidt modes $f_k$ with weights $w_k$ and
the local oscillator sits in mode $g$, each component contributes an
interference term weighted by $|\langle g|f_k\rangle|^2$, so $\xi^2=\sum_k
w_k|\langle g|f_k\rangle|^2$; Eq.~(\ref{eq:xieff}) is the case in which $g$
overlaps $f_1$ alone. We verified both forms to $6\times10^{-16}$ against a
direct construction of the mixture, for either detector and for the marginals
as well as the joint probability. Section~\ref{sec:exp} returns to the
consequence.

\emph{Phase noise.} The correlator depends on $\alpha$ and $\beta$ only
through $|\alpha|^2$, $|\beta|^2$ and $\mathrm{Re}[(\alpha^*\beta)^N]$. A
phase drift \emph{common} to the two stations leaves all three invariant and
cancels identically. Only the differential phase matters, and taking it
Gaussian with rms $\sigma$ damps the interference term by
$e^{-N^2\sigma^2/2}$. This is the formal counterpart of the design of
Guerreiro \textit{et al.}\ and Caspar \textit{et al.}, who co-propagate the
local oscillator with the signal so phase fluctuations are common-mode and no
active lock is needed \cite{Guerreiro16,Caspar22}.

\emph{Upstream loss.} Let $p$ be the probability that a heralded photon
arrives in the two signal modes, collecting heralding efficiency, fibre
coupling and every optical loss ahead of the displacement. Heralding failure
is equivalent to loss on the signal modes only in the low-gain limit, where
higher-order pairs are negligible; we return to that in Sec.~\ref{sec:exp}.
For a single photon this is exactly a vacuum admixture,
\begin{equation}
  \rho(p)=p\,|\Psi\rangle\langle\Psi|+(1-p)\,|0,0\rangle\langle0,0|.
  \label{eq:rhop}
\end{equation}
The Kraus element in which no photon is lost acts as $\sqrt p$ times the
identity on the one-photon sector. The two elements that do lose the photon
both land on $|0,0\rangle$, but they carry different Kraus indices, so they
add incoherently; coherently they would cancel exactly.
Equation~(\ref{eq:rhop}) assumes equal loss in the two arms; with $p_a\neq
p_b$ the state is partially unbalanced rather than a pure vacuum admixture,
and the thresholds below are for the symmetric case, a restriction
Sec.~\ref{sec:invariance} returns to.

\subsection{The single-photon correlator}

For $N=1$ everything collapses into one expression. With $q=e^{-\sigma^2/2}$,
\begin{align}
  \Pi_x(\alpha,\beta)={}&e^{-4\nu}e^{-\lambda(|\alpha|^2+|\beta|^2)}
   \Big[1-p\lambda\nonumber\\
  &+\tfrac12 p\lambda^2\xi^2\big(|\alpha|^2+|\beta|^2
   -2q\,\mathrm{Re}(\alpha^*\beta)\big)\Big].
  \label{eq:Pi1}
\end{align}
Equation~(\ref{eq:Pi1}) is Eq.~(\ref{eq:Pigen}) at $N=1$ with the
imperfections included; for parity detection $E=\Pi_x$ directly. The on--off
scheme also needs the single-detector terms, since it assigns $+1$ to no click
and $-1$ to a click, giving $\langle A\rangle=2\langle\hat M\rangle-1$. Write
$P(\alpha,\beta)$ for the joint probability that neither detector clicks, and
$P(\alpha)$ for the corresponding marginal at one station, the same symbols as
in Eq.~(\ref{eq:ch}). The former is Eq.~(\ref{eq:Pi1}) with the $e^{-4\nu}$
factor removed; it is the two-mode $Q$ function only at $\eta=1$, and
otherwise the $s$-ordered function of Sec.~\ref{sec:family}. Then
\begin{align}
  P(\alpha)&=e^{-\lambda|\alpha|^2}
    \Big[1-\tfrac12 p\lambda+\tfrac12 p\lambda^2\xi^2|\alpha|^2\Big],
  \label{eq:Q1}\\
  E_{\text{on-off}}(\alpha,\beta)&=4e^{-2\nu}P(\alpha,\beta)\nonumber\\
    &\quad-2e^{-\nu}P(\alpha)-2e^{-\nu}P(\beta)+1.
  \label{eq:Eonoff}
\end{align}
Every imperfection now appears once, and in a different place. Dark counts
appear as an overall factor for parity and through $e^{-\nu}$ for on--off, and
upstream loss as a weight on the signal terms. Mode mismatch multiplies the
whole $p\lambda^2$ group, interference and single-mode terms alike, because an
imperfectly matched local oscillator rescales the displacement seen by the
signal; phase noise acts on the interference term alone.

Summing Eq.~(\ref{eq:Eonoff}) over the four CHSH terms, the marginals at the
primed settings cancel, those at the unprimed settings add, and the four
constants leave $+2$, so that
\begin{align}
  S-2=4\Big[&P(\alpha,\beta)+P(\alpha',\beta)+P(\alpha,\beta')\nonumber\\
   &-P(\alpha',\beta')-P(\alpha)-P(\beta)\Big]
  \label{eq:CHident}
\end{align}
at $\nu=0$. The bracket is exactly Eq.~(\ref{eq:ch}). This standard
equivalence of the two inequalities for dichotomic observables holds for any
dichotomic observable $\hat A=2\hat M-1$ built from the effect $\hat M$ of a
two-outcome measurement. $\hat M$ need not be a projector: the no-click effect
is one only at $\eta=1$, so the on--off scheme is exactly the non-projective
case. Here the identity implies that the on--off scheme runs the
Clauser--Horne test of Ref.~\cite{Wildfeuer07} in a different notation. A CHSH
violation, $|S|>2$, is a Clauser--Horne violation, $\mathrm{CH}\notin[-1,0]$,
and conversely, so the on--off results below add no new inequality to the
Clauser--Horne analysis already published there. What they do add is its
dependence on efficiency and noise, which Ref.~\cite{Wildfeuer07} treated only
for ideal detectors. The two-sidedness matters in practice: the large
violation for this state sits on the branch $\mathrm{CH}<-1$, and a numerical
search that looks only for $\mathrm{CH}>0$ finds a much weaker optimum.

\section{Only the overall efficiency matters}
\label{sec:invariance}

Upstream loss and detector loss do not trade off in any complicated way, and
the reason needs no closed form. The loss channel is displacement covariant,
\begin{equation}
  \Phi_\eta\big[\hat D(\alpha)\rho\hat D^\dagger(\alpha)\big]
  =\hat D(\sqrt\eta\,\alpha)\,\Phi_\eta[\rho]\,\hat D^\dagger(\sqrt\eta\,\alpha),
  \label{eq:cov}
\end{equation}
and loss channels compose, $\Phi_p\circ\Phi_\eta=\Phi_{p\eta}$. The physical
sequence --- upstream loss $p$, displacement $\alpha$, detector loss $\eta$,
ideal detector --- is therefore the same POVM on the source state as upstream
loss $p\eta$, displacement $\sqrt\eta\,\alpha$, ideal detector. Since the
displacement amplitudes are optimised over, the rescaling is free, and we have

\begin{quote}
\emph{With symmetric arms, the maximum Bell value depends on upstream loss and
detector inefficiency only through the product $\mu=p\eta$.}
\end{quote}

The statement holds for any $N$, any input state and either detector, and with
dark counts, mode mismatch and phase noise switched on. Three hypotheses are
doing work and should be named. The loss must act identically on the two arms;
the displacement amplitude must be unconstrained, since the rescaling by
$1/\sqrt\eta$ is what makes it free, so a modulator of finite range would
break the collapse at small $\eta$; and the three imperfections must commute
with that rescaling. Dark counts do because they act at the ideal detector,
after all loss. Phase noise does because it touches only relative phases. Mode
mismatch is the one that is not obvious: it commutes because the unmatched
local-oscillator component passes through the same detector loss as the
matched one, so $\xi$ enters only in the combination $\xi\alpha$ and the whole
$\alpha$-dependence rescales together. We verified Eq.~(\ref{eq:cov}) and the
resulting correlator invariance to $8\times10^{-16}$ by explicit
Kraus-operator computation in a truncated Fock space, using no analytic
expression. That coupling, transmission and detection can be collected into a
single efficiency is used without comment by Brask and Chaves
\cite{BC12,CB11}.

Symmetry is needed only for the last step. The covariance argument gives a
separate dependence on $\mu_a=p_a\eta_a$ and $\mu_b=p_b\eta_b$; collapsing the
two into one number requires $\mu_a=\mu_b$. Numerically the collapse is
benign: the threshold is very nearly a threshold on the geometric mean
$\sqrt{\mu_a\mu_b}$. Writing the imbalance as the ratio $\mu_a/\mu_b$, that
threshold falls from the $\xi=0.97$ value $0.8419$ of
Sec.~\ref{sec:thresholds} to $0.8412$ at $\mu_a/\mu_b=0.9$, and to $0.8387$ at
$0.8$, so imbalance is mildly favourable rather than costly.

For $N=1$ there is a second, more elementary route to the theorem: upstream
loss is then exactly the vacuum admixture of Eq.~(\ref{eq:rhop}), and
substituting $\alpha\to\tilde\alpha/\sqrt\eta$ in Eq.~(\ref{eq:Pi1}) leaves
$p$ and $\eta$ only in the combination $\mu$. That route does not generalise,
but Eq.~(\ref{eq:cov}) does. From here on, settings are quoted in the
representation used above, with all loss carried by the state and the detector
ideal; the displacement a physical apparatus must apply is larger by
$1/\sqrt\eta$. The amplitudes quoted in Sec.~\ref{sec:family}, where the
detector carries its own loss, are already physical.

\begin{figure}[t]
  \includegraphics[width=\columnwidth]{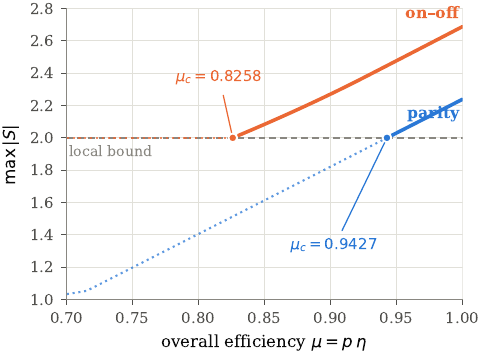}
  \caption{Maximum CHSH value for the state of Eq.~(\ref{eq:psi1}) against
    overall efficiency $\mu=p\eta$, with no dark counts, perfect mode overlap
    and no phase noise. Dotted below threshold. The on--off curve saturates at
    the local bound rather than falling through it because a large local
    oscillator makes that detector click deterministically, reproducing
    $|S|=2$. Parity has no such trivial setting for $0<\mu<1$, and its curve
    therefore falls below the local bound.}
  \label{fig:mu}
\end{figure}

\section{Thresholds}
\label{sec:thresholds}

\subsection{Optimal settings}

The four settings are complex, so the optimisation is over eight real
parameters, reduced to seven by the global phase. We searched with
differential evolution over $[-3.5,3.5]^8$ from several seeds, followed by a
few hundred Nelder--Mead restarts and by continuation in $\mu$; no functional
form was assumed. This matters, because a natural-looking reduction fails.

At $\mu=1$ the optimum is antisymmetric and collinear: $\beta=-\alpha$,
$\beta'=-\alpha'$, with $\alpha$ and $\alpha'$ real and antiparallel. Parity
detection gives $\max|S|=2.2387$ at $|\alpha|=0.0610$, $|\alpha'|=0.3391$,
consistent with Ref.~\cite{Wildfeuer07}; on--off detection, with $+1$ assigned
to no click, gives $\max|S|=2.6884$ at $|\alpha|=0.1653$, $|\alpha'|=0.5630$,
in agreement with Lee \textit{et al.}\ \cite{Lee17}. It is tempting to impose
that structure and reduce the problem to two parameters. For parity detection
that is safe wherever the test is of interest: we find the collinear
configuration optimal for every $\mu\ge0.72$, which contains the whole
violating region with room to spare. It fails below that --- the $x=0$ optimum
quoted in Sec.~\ref{sec:family}, with its $\pm135^\circ$ phases, is what the
true optimum looks like at $\mu=1/2$, and it is not collinear --- but there is
no violation left there to lose. For on--off detection the reduction is not
safe even near threshold. Below $\mu\approx0.9$ the optimum acquires
non-trivial relative phases,
\begin{align}
  \mu=0.85:\quad
  &|\alpha|=|\beta|=0.374, \quad \arg(\alpha^*\beta)=-62.6^\circ,\nonumber\\
  &|\alpha'|=|\beta'|=0.630, \quad \arg(\alpha'^*\beta')=+148.5^\circ,\nonumber\\
  &\arg(\alpha^*\beta')=\arg(\alpha'^*\beta)=+42.9^\circ,
  \label{eq:optset}
\end{align}
and restricting the settings to real numbers discards them.
Equation~(\ref{eq:optset}) gives all of them: setting four modulators takes
four magnitudes and three independent relative phases, the fourth pairwise
argument being fixed by the other three. The third line holds because this
solution is invariant under exchanging the two stations together with complex
conjugation, which is not automatic. The magnitudes remain equal across the
two stations, but the phases do not lock to $0$ or $\pi$. The difference is
not small near threshold: at $\mu=0.83$ the collinear family gives no
violation at all, while the true optimum gives $|S|=2.0143$. The
single-detector terms of Eq.~(\ref{eq:Eonoff}) break the symmetry; the parity
correlator has none, so only the on--off branch is affected.

Neither assignment discards data (Sec.~\ref{sec:debate}), so the efficiencies
below are thresholds for a test free of the detection loophole, not for a
post-selected one.

\subsection{Threshold efficiencies}

Figure~\ref{fig:mu} shows the degradation. Thresholds are located by bisecting
the predicate $\max|S|>2+\epsilon$ with $\epsilon=10^{-9}$, rather than by
root finding, because for on--off detection $|S|=2$ is a supremum attained as
$|\alpha|\to\infty$, where the detector clicks deterministically. We find
\begin{equation}
  \mu_c^{\,\text{on-off}}=0.8258,
  \qquad
  \mu_c^{\,\text{parity}}=0.9427 .
  \label{eq:muc}
\end{equation}
The first is not new. It agrees with the $0.825$ overall efficiency of Lee
\textit{et al.}\ for this scheme \cite{Lee17}, with the $s=-1$ threshold of
Lee, Jeong and Jaksch \cite{LJJ09}, and with the $82.6\%$ quoted by Alwehaibi
\textit{et al.}\ for this state \cite{Alwehaibi25}. The parity value we have
not found in the literature. It sits below the $96\%$ that D'Angelo \textit{et
al.}\ quote for a loophole-free test of the Banaszek--W\'odkiewicz inequality
\cite{DAngelo06}, the difference being that Eq.~(\ref{eq:muc}) optimises
freely over all four settings whereas that inequality fixes them to a
one-parameter family.

On--off detection, which needs no photon-number resolution at all, is better
on both counts: it violates more strongly at unit efficiency and tolerates
twelve percentage points more loss. The first fact is already in the
literature \cite{Lee17}; the second decides an experiment, and
Eq.~(\ref{eq:lossparity}) says why. The detection-efficiency threshold for a
maximally entangled pair of qubits under the same inequality, with
non-detection assigned a fixed outcome, is $2(\sqrt2-1)=0.8284$
\cite{GargMermin87}, so the displacement test on a single delocalised photon
is marginally \emph{easier} than the textbook case.
Section~\ref{sec:practical} finds that unbalancing this state raises the
threshold rather than lowering it. Together those two facts fix the character
of the problem: this is a standard-difficulty Clauser--Horne--Shimony--Holt
test, on which Eberhard's trick of weakening the state --- which buys so much
for polarisation pairs --- is unavailable. Equation~(\ref{eq:lossparity}) is
the reason: the alternating unit-magnitude weight $(-1)^{\hat n}$ that makes
parity a sharp probe of nonclassicality becomes $(1-2\eta)^{\hat n}$, and is
the first thing loss destroys, while the on--off weight $(1-\eta)^{\hat n}$
degrades far more slowly with $\eta$.

Neither threshold is the best available. Adding a squeezer before the on--off
detector lowers it to $\mu\approx0.78$ \cite{Lee17}, and the efficiency a
photon-pair Bell test needs is optimised over measurement strategies in
Ref.~\cite{Vivoli15}. Protocols that herald a tunable superposition of vacuum
with $|1,1\rangle$, rather than distributing a single photon, approach the
Eberhard limit of $2/3$ \cite{BC12,Alwehaibi25}. The general point --- that
weakly entangled states resist the detection loophole better --- is due to
Eberhard \cite{Eberhard93} and is developed for these states in
Refs.~\cite{CB11,BC12}. Long-distance variants of the displacement measurement
itself are analysed in Refs.~\cite{Mycroft23,Bjerrum23}.

\begin{figure*}[t]
  \includegraphics[width=\textwidth]{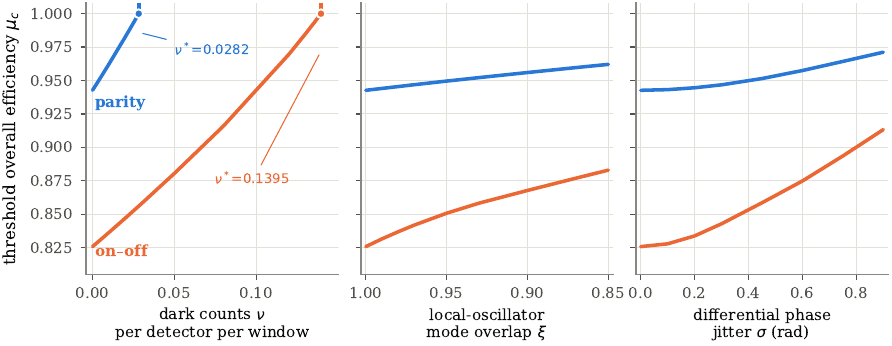}
  \caption{Threshold overall efficiency for a CHSH violation with $N=1$ as each
    remaining imperfection is switched on, from the full complex search. The
    dark-count curves rise vertically at $\nu^*=0.0282$ for parity, the closed
    form of Eq.~(\ref{eq:nustar}), and at $0.1395$ for on--off, found
    numerically; beyond these no efficiency suffices. Note the different
    horizontal scales: mode overlap runs right to left, so that all three
    panels degrade to the right.}
  \label{fig:noise}
\end{figure*}

\subsection{Dark counts}

Dark counts are the hard wall. Because of the exact scaling
Eq.~(\ref{eq:dark}), the parity violation vanishes at
\begin{equation}
  \nu^*_{\text{parity}}=\tfrac14\ln\big(\max|S|_{\mu=1}/2\big)=0.0282
  \label{eq:nustar}
\end{equation}
counts per detector per window, and no efficiency, however high, recovers it.
The on--off scheme, whose dark-count dependence is not a rescaling, survives
to $\nu^*_{\text{on-off}}=0.1395$, a factor of $4.95$ more. At a modest
$\nu=0.01$ the parity threshold has already climbed from $0.943$ to $0.962$
while the on--off threshold moves only from $0.826$ to $0.836$
(Fig.~\ref{fig:noise}).

These are enormous numbers by the standards of superconducting detectors. With
timing jitter from below $3$~ps to a few tens of ps
\cite{Korzh20,EsmaeilZadeh20,Chang21} a coincidence window of $100$~ps is
comfortable, and at even $500$ counts per second, the value \emph{measured}
rather than projected in Ref.~\cite{Chang21}, that is $\nu\sim5\times10^{-8}$.
Dark counts are five to six orders of magnitude from mattering here.

\subsection{Mode matching and phase stability}
\label{sec:noise}

Neither is the interferometry, although it is not quite free. We read the
indistinguishability reported in Ref.~\cite{Guerreiro16} as an
\emph{amplitude} overlap, the conservative reading; taken as an intensity
overlap it would give $\xi=0.985$ and a threshold of $0.834$. A local
oscillator with $\xi=0.97$ raises the on--off threshold from $0.8258$ to
$0.8419$ and the parity threshold from $0.9427$ to $0.9469$. That is $1.6$
percentage points for the on--off scheme, not the negligible amount one might
guess. At $\xi=0.85$ the two become $0.883$ and $0.962$. Differential phase
jitter is gentler: $\sigma=0.3$~rad, some $17^\circ$ rms, costs the on--off
scheme less than two percentage points, and even $\sigma=0.9$~rad leaves both
tests alive. Actively stabilised fibre interferometers hold the relative phase
to about $1$~mrad over hours \cite{Svarc23}. Co-propagation helps but does not
remove the need for an active lock. It cancels the \emph{common path length},
which Sec.~\ref{sec:corr} shows drops out exactly. But signal and local
oscillator are carried in orthogonal polarisations, and the birefringent phase
between them is not common-mode. Over tens of metres that term dominates.
Accordingly Caspar \textit{et al.}, distributing this state over distance,
used a piezoelectric fibre stretcher, a counter-propagating reference laser
with feedback, and electronic polarisation control \cite{Caspar22}. None of
that costs signal loss provided it acts on the local-oscillator side, but it
must be in the design. The experimental budget of Sec.~\ref{sec:exp} is
therefore measured against $\mu_c=0.8419$, the $\xi=0.97$ value; elsewhere we
quote the $\xi=1$ thresholds, which isolate the effect under study.

\section{What an experiment needs}
\label{sec:exp}

\begin{table*}[t]
\caption{Best demonstrated values for the quantities entering $\mu=p\eta$ and
the noise parameters. Two cautions. The $810$~nm heralding record and the
telecom detector record cannot be combined, although Ref.~\cite{Liu22}
supplies both at $1560$~nm. And the coupling of Ref.~\cite{Liu22} and the overlap of
Ref.~\cite{Guerreiro16} are two halves of one trade-off
(Sec.~\ref{sec:budget}).}
\label{tab:hardware}
\begin{ruledtabular}
\begin{tabular}{llc}
quantity & best demonstrated & Ref. \\ \hline
SNSPD efficiency, 1550~nm         & $0.980\pm0.005$          & \cite{Reddy20} \\
SNSPD efficiency, 850~nm          & $0.76$                   & \cite{Los24} \\
SNSPD efficiency, 940~nm          & $0.945$                  & \cite{Los24} \\
number-resolving SNSPD, 1555~nm   & $0.98$, resolves $n\le32$ & \cite{PNRsnspd25} \\
TES efficiency, 1550~nm           & $0.98$, microsecond recovery  & \cite{Fukuda11} \\
SPDC heralding into SMF, 810~nm   & $0.844\pm0.005$          & \cite{Pereira13} \\
SPDC coupling into SMF, 1560~nm & $0.956$ & \cite{Liu22} \\ SPDC heralding
into SMF, 1541~nm & $\approx0.80$ & \cite{Guerreiro16} \\ system efficiency
$\mu$, telecom & $0.872\pm0.002$ & \cite{Liu22} \\ system efficiency $\mu$,
$810$~nm & $0.786$ / $0.762$ & \cite{Giustina15} \\
quantum dot, at fibre output      & $0.712\pm0.018$          & \cite{Ding25} \\
photon vs.\ coherent overlap $\xi$ & $>0.97$                 & \cite{Guerreiro16} \\
phase stability $\sigma$          & $\sim1$~mrad over hours  & \cite{Svarc23} \\
dark counts, measured / filtered  & $300$--$500$ / $1$--$10$~s$^{-1}$ & \cite{Chang21} \\
\end{tabular}
\end{ruledtabular}
\end{table*}

Section~\ref{sec:invariance} reduces the requirement to a single number.
Table~\ref{tab:hardware} collects the best each factor has reached. The
experiment we have in mind is the one already built in Geneva
\cite{Monteiro15,Guerreiro16,Caspar22}. A heralded photon from parametric
down-conversion is split on a fibre beam splitter; each arm is displaced by
mixing with a co-propagating local oscillator; each output goes to a
superconducting nanowire detector.

\subsection{Efficiency is not the obstacle; overlap is}
\label{sec:budget}

Efficiency alone is no longer the obstacle. Liu \textit{et al.}\ report, in a
single telecom apparatus, a fibre coupling of $0.956$, an optical transmission
of $0.935$ and a detector efficiency of $0.975$, for a heralded detection
efficiency of $0.872$ \cite{Liu22}. Their optics are a polarisation analyser,
which a displacement station would replace. Take their coupling with this
paper's detector and interferometer, and write $T_{\text{opt}}$ for the
transmission from the collection fibre to the detector. Then
\begin{equation}
  \mu=0.956\times T_{\text{opt}}\times0.980 ,
  \label{eq:mutel}
\end{equation}
which is $0.862$ at the buildable $T_{\text{opt}}=0.92$ and $0.909$ at an
optimistic $0.97$. Both exceed $\mu_c=0.8419$. At $810$~nm the gap is already
small without any construction at all. The system efficiency Giustina
\textit{et al.}\ \emph{measured} --- two-fold coincidences over singles,
uncorrected for anything --- is $0.786$ and $0.762$ for their two arms
\cite{Giustina15}. That is $\mu$ itself, six and eight points from threshold,
in an apparatus not built for this.

Multiplying the records of Table~\ref{tab:hardware} therefore gives the wrong
answer, and gives it in the pessimistic direction. What it misses is
Eq.~(\ref{eq:xieff}): the overlap $\xi$ that sets $\mu_c$ carries a factor
$\sqrt{w}$, and $w$ is not free. A heralded photon is spectrally pure only if
the two-photon state is close to factorable, and for a source with the
frequency correlation a narrowband pump produces, making it factorable means
filtering away exactly the pairs that heralding efficiency counts
\cite{MeyerScott17}. The coupling of Ref.~\cite{Liu22} and the overlap of
Ref.~\cite{Guerreiro16} are thus two ends of one trade-off, reached in
different laboratories. Liu \textit{et al.}\ used a $10$~ns, effectively
quasi-continuous pump and report neither purity nor overlap; Guerreiro
\textit{et al.}\ reported an overlap above $0.97$ at a heralding efficiency of
about $0.80$. No device reports both, and none reports $w$.

\begin{figure}[t]
  \includegraphics[width=\columnwidth]{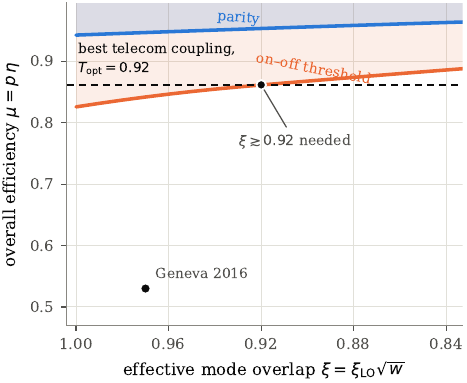}
  \caption{The frontier the experiment has to cross. Shaded regions violate,
    on--off below and parity above. The horizontal axis is the effective
    overlap of Eq.~(\ref{eq:xieff}), which carries the heralded photon's
    single-mode weight as well as the local oscillator's mode match; it runs
    right to left, so both axes degrade downward and to the right. The dashed
    line is the overall efficiency reachable with the best demonstrated telecom
    coupling \cite{Liu22} and a buildable station, $\mu=0.862$, and it crosses
    the on--off contour at $\xi=0.92$. The Geneva apparatus \cite{Guerreiro16}
    is plotted at its own reported efficiency and overlap. }
  \label{fig:frontier}
\end{figure}

Figure~\ref{fig:frontier} is the resulting picture. At Liu \textit{et al.}'s
coupling with a buildable station the threshold contour crosses $\mu=0.862$ at
an effective overlap $\xi=0.92$ --- numerically equal to $T_{\text{opt}}$
above, but an unrelated quantity --- which by Eq.~(\ref{eq:xieff}) is a
single-mode weight of at least $0.85$ against a perfect local oscillator. We
read Guerreiro \textit{et al.}'s reported indistinguishability as an
\emph{effective} overlap, since it is an interference contrast and whatever
single-mode weight their source had is already in it; that puts them above the
requirement on one axis and far below it on the other. A source engineered for
group-velocity matching avoids the trade-off rather than navigating it, and
the figure of merit to report is the pair $(h,\,\xi)$ measured on one device
--- which no source paper at present does. The Geneva apparatus itself sits
far below both contours. Guerreiro \textit{et al.}\ report detector $0.85$,
fibre coupling $0.80$, optical transmission $0.78$ and heralding about $0.80$,
and quote $52\%$ as the probability of detecting the heralded photon, which is
the first three factors to within a point. Whether their heralding and
coupling efficiencies are the same quantity described twice, as the Klyshko
definition would make them, is not stated; taking them to be the same gives
$\mu_{\text{Geneva}}=0.53$, and reading them as independent gives $0.42$. We
quote the larger, being the reading less favourable to the argument here.

\subsection{What the station may cost}
\label{sec:station}

The efficiency budget is best inverted, and best stated in decibels, because
that is the currency fibre components are specified in. Reaching
$\mu_c=0.8419$ at $\eta=0.980$ requires $h\,T_{\text{opt}}\ge0.859$, so the
entire chain from the collection fibre to the detector pigtail may lose no
more than Table~\ref{tab:station} allows.

\begin{table}[b]
\caption{Loss the displacement station may spend, as a function of the
coupling efficiency $h$ of the source, at $\eta=0.980$ and the on--off
threshold $\mu_c=0.8419$. Below $h=0.859$ no station is good enough.}
\label{tab:station}
\begin{ruledtabular}
\begin{tabular}{ll}
coupling $h$ & allowed station loss \\ \hline
$0.956$ \cite{Liu22}    & $0.46$~dB \\
$0.90$                  & $0.20$~dB \\
$0.886$                 & $0.13$~dB \\
$0.85$                  & unreachable at any loss \\
\end{tabular}
\end{ruledtabular}
\end{table}

Half a decibel is not much for a chain that must contain a fibre beam
splitter, polarisation control, the displacement coupler, splices and the
detector pigtail. The design rule of Sec.~\ref{sec:systematics} matters here
too: every element moved into the local-oscillator arm is an element removed
from this budget. In particular a fixed fused tap should replace the Geneva
design's variable-ratio coupler, with the setting applied to the local
oscillator upstream, which both removes the setting-dependent transmission and
returns the coupler's insertion loss to the signal path.

\subsection{Statistics and systematics}
\label{sec:systematics}

The violation is not hard to accumulate. All numbers in this subsection are
for the realistic case $\xi=0.97$. At $\mu=0.86$, where $|S|=2.067$, the four
optimal correlators are $(-0.56,-0.56,-0.56,+0.38)$, giving $\sum(1-E^2)=2.91$
and $\sigma_S=3.41/\sqrt{R}$, with $\sigma_S$ the standard error of $S$ and
$R$ the number of trials split evenly over the four setting pairs. A
five-standard-deviation violation therefore needs $R\approx6.6\times10^{4}$:
half a minute at the $2$~kHz herald rate of Ref.~\cite{Guerreiro16}, six
seconds at the $11.5$~kHz of Ref.~\cite{Caspar22}.

The even split is a variance estimate, not a protocol: the settings must be
drawn independently at random at each station on every trial, since a
balanced-by-construction allocation is exactly what a device with memory
exploits. And standard deviations are the wrong currency for a loophole-free
claim: if the devices may have memory the trials are not independent, so the
central limit theorem supplies no null distribution to put error bars on. What
a bounded per-trial estimator does supply is a $p$-value valid against
arbitrary memory \cite{BCHKM02,Gill03}, with $R$ fixed in advance and the
setting generator independent of the past. With uniform random settings that
estimator lies in $[-4,4]$, so Azuma--Hoeffding gives
$p\le\exp(-R\delta^2/32)$ with $\delta=|S|-2$; a $p$ of $2.9\times10^{-7}$,
the Gaussian five-sigma equivalent, then needs $R=1.1\times10^{5}$ --- a
factor $1.7$ above the estimate that assumed independent trials. Sharper
bounds do better \cite{Bierhorst15}. The two inequalities, though
algebraically equivalent by Eq.~(\ref{eq:CHident}), are not equivalent as
statistical tests on finite data \cite{Renou17}, so the choice between them is
a question about the analysis rather than about the apparatus.

The experiment is therefore systematics-limited, and two effects must be kept
apart. Neither is a loophole, for the reason given in Sec.~\ref{sec:debate}. A
\emph{uniform} efficiency drift changes the true $S$ at the rate
$\mathrm{d}S/\mathrm{d}\mu=3.76$, so a relative drift of $0.3\%$ moves $S$ by
$0.010$, a seventh of the violation, as often downward as upward. A
\emph{setting-dependent} transmission is worse in two ways: it is larger, and
it has a definite sign. Taking the transmission to differ by $\pm d$ between
the two settings at each station, we find $\Delta S=+0.0027$ at $d=0.3\%$,
$+0.0091$ at $d=1\%$, and the whole violation at $d\approx7.5\%$. The sign is
set by the mechanism, not by anything general --- reversing the correlation
between setting and transmission reverses it --- but here it is the inflating
one, so the apparatus reports a better violation than it has. The diagnostic
is to compare the measured single-station marginals, setting by setting,
against Eq.~(\ref{eq:Q1}). In the Geneva design the displacement is a
polarisation controller followed by an in-line polariser, so changing the
setting changes the projection angle and hence the \emph{signal} transmission,
and the two settings here are $|\alpha|=0.30$ and $0.61$. A coupler of
transmissivity $T_d$ fed with a fixed local oscillator of $|\gamma|^2$ photons
transmits the signal with probability $1-|\alpha|^2/|\gamma|^2$, so the
setting-dependent part is $|\alpha|^2/|\gamma|^2$: $0.07\%$ at the
$|\gamma|^2\approx400$ below, worth $\Delta S=+0.0003$, but $0.3\%$ and
$+0.0013$ at the hundred photons per pulse the Geneva apparatus uses. One
design rule removes it altogether: hold the coupling ratio fixed and vary the
amplitude in the local-oscillator arm alone, upstream of the displacement
coupler but inside the station, where a modulator's insertion loss never
touches the signal. In a co-propagating layout that is awkward: signal and
local oscillator share one fibre, so acting on the latter alone needs a
polarisation demultiplex and recombination, costing several tenths of a
decibel of \emph{signal} loss --- most of the station budget of
Sec.~\ref{sec:station}. The alternative is to give the local oscillator its
own fibre, at the price of the phase stability that Sec.~\ref{sec:noise}
argues must be actively bought in any case. The optimum also moves in phase as
well as amplitude --- $\arg(\alpha^*\beta)=-77^\circ$ at one setting pair
against $\arg(\alpha'^*\beta')=+156^\circ$ at the other --- so each station
needs a fast local modulator of both.

Dead time is the one item here that \emph{is} a loophole. A detector still
recovering cannot click, so the trial has one of two fates. Either it goes
unrecorded, which breaks the first condition of Sec.~\ref{sec:debate}: the
surviving ensemble has then been selected by the setting, and that is a
genuine loophole. Or it is recorded as a forced $+1$, not a loophole but a
bias no efficiency budget captures. Either way it is the one mechanism in the
apparatus that correlates a trial with its predecessor, which is precisely the
memory the bound above exists to defend against.

The size of that effect forces the local oscillator to be gated to the
heralded windows. At the two settings, $|\alpha|=0.30$ and $0.61$, the local
oscillator alone makes a detector click with probability $0.09$ and $0.31$, in
every window in which it is present. Left free-running at a $9.5$~MHz pump
rate, that is $8\times10^5$ and $3\times10^6$ counts per second, and at a
$50$~ns recovery time the dead fraction is $4\%$ at one setting and $15\%$ at
the other. Gated to an $11.5$~kHz herald rate the same fractions are
$5\times10^{-5}$ and $2\times10^{-4}$; treated as an efficiency reduction that
is worth $\Delta S=-3\times10^{-4}$, and even a transition-edge sensor with
microsecond recovery gives only $-0.006$. Gating is therefore a requirement,
not an optimisation, and it brings a timing constraint of its own. With the
source midway, the signal photon reaches a station in $D n/2c=142$~ns at
$D=58$~m, while the herald needs detector and discriminator latency plus the
same $29$~m of cable, which is comparable. The gate must therefore either sit
at the source, before the long fibre, or be fed by a herald that arrives in
time, with a short optical delay in the signal arm to buy the margin. Its rise
time must be electro-optic; an acousto-optic modulator is far too slow.

That local oscillator is weak. The physical displacement is
$|\alpha|/\sqrt\eta$, so with $|\alpha|=0.61$ and $\eta=0.98$ one needs
$|\gamma|^2=|\alpha|^2/[\eta(1-T_d)]$ photons per pulse: about $400$ at
$1-T_d=10^{-3}$, a peak power of tens of nanowatts in a nanosecond pulse and,
gated to the herald rate, a mean power of picowatts. The peak is four to five
orders below conventional homodyne practice, the mean nine or ten. It
constrains which modulators are usable, and it means the displacement must be
calibrated on the pulse peak or by photon counting rather than on an
average-power photodiode.

\subsection{Multiphoton contamination}
\label{sec:multiphoton}

Equation~(\ref{eq:rhop}) is exact for photon loss but not for higher-order
down-conversion pairs, the one error channel of a heralded source that is not
loss. We treat them exactly.

Take a two-mode squeezed source of pair probability $P_{\text{pair}}$ per
pulse, heralded by a bucket detector of efficiency $\eta_h$. The signal is
left in the Fock mixture $w_k\propto P_{\text{pair}}^{\,k}[1-(1-\eta_h)^{k}]$,
whose two-photon fraction is
\begin{equation}
  f_2\equiv\frac{w_2}{w_1}=P_{\text{pair}}(2-\eta_h).
  \label{eq:f2}
\end{equation}
A bucket herald fires on two pairs more readily than on one, so heralding is
\emph{biased towards} the contamination, and the bias is worst for a poor
herald detector. After the beam splitter the two-photon term is
$(|2,0\rangle-\sqrt2|1,1\rangle+|0,2\rangle)/2$, so it contributes doubly
occupied arms as well as the $|1,1\rangle$ component. A vacuum admixture can
produce neither, and the doubly occupied arms matter because the observable is
\emph{no click}.

Carrying this through the loss channel and the on--off detection involves no
approximation: loss cannot raise the photon number, so the state keeps bounded
support and the coherent-state overlaps are analytic. Optimising over all four
complex settings as before, we find that the threshold shift depends on
$P_{\text{pair}}$ and $\eta_h$ only through $f_2$, and that
\begin{equation}
  \Delta\mu_c\simeq0.23\,f_2
  \label{eq:multi}
\end{equation}
to within $2\%$ for $f_2\lesssim10^{-2}$ and $9\%$ out to
$f_2=4.5\times10^{-2}$, the largest contamination we computed. The collapse
onto $f_2$ alone is exact only within the two-pair truncation: the three-pair
weight $w_3/w_1=P_{\text{pair}}^2(3-3\eta_h+\eta_h^2)$ is not a function of
$f_2$, and including it leaves a residual dependence on $\eta_h$ at fixed
$f_2$ of under $2\%$ of $\Delta\mu_c$ across the range we computed. A source
of Schmidt number $K$ contaminates less at the same mean pair number, by a
factor $(K+1)/2K$, so the single-mode $f_2$ is conservative --- but the same
$K$ enters Eq.~(\ref{eq:xieff}) as a reduced weight $w$, where it costs far
more than it saves.

The penalty is therefore negligible where these experiments operate. Caspar
\textit{et al.}\ quote a two-photon fraction of $5\times10^{-3}$
\cite{Caspar22}, giving $\Delta\mu_c=1.2\times10^{-3}$, an order of magnitude
below the $1.6\times10^{-2}$ that imperfect mode matching costs. Contamination
becomes as expensive as mode matching only near $f_2\approx0.06$, where the
effective coefficient has risen to about $0.26$ and Eq.~(\ref{eq:multi}) is no
longer a good fit; that is a per-pulse pair probability of a few per cent.

The design consequence is in the $2-\eta_h$. The herald rate scales as
$P_{\text{pair}}\eta_h$, so buying rate by raising the pump raises $f_2$ in
proportion, while buying it by improving the herald detector \emph{lowers}
$f_2$. Only one of those two moves is free. Here $\eta_h$ is the efficiency of
the detector on the \emph{idler}. It is not the upstream efficiency $p$ of
Eq.~(\ref{eq:rhop}), which is a property of the signal arm and of which
heralding is only one factor. In the low-gain limit the two are independent,
so improving the herald detector buys rate and purity but does not move $\mu$.

\subsection{Outside the model}
\label{sec:outside}

Section~\ref{sec:debate} makes two spacelike-separation conditions of the
test, and they are not equally forgiving. Locality requires that each setting
choice be generated, applied, and the outcome recorded, before light could
carry news of it to the far station:
\begin{equation}
  \tau_{\text{RNG}}+\tau_{\text{mod}}+W\le D/c,
  \label{eq:timing}
\end{equation}
with $D$ the station separation and $W$ the detection window. Measurement
independence requires more. With the source midway, each choice must also be
spacelike separated from the emission, which costs a further factor of two,
\begin{equation}
  \tau_{\text{RNG}}+\tau_{\text{mod}}\le D/2c,
  \label{eq:free}
\end{equation}
with one consequence that is easy to miss. A setting choice \emph{triggered by
the herald} lies inside the emission light cone by construction, so the
settings must run on a free clock and the herald must serve only to define the
trial. At the $58$~m of Ref.~\cite{Giustina15} the two budgets are $193$~ns
and $97$~ns; the second binds, and their $26$~ns of setting generation leaves
some $71$~ns for a modulator that must move both amplitude and phase ---
tight. Both budgets have been met in photonic tests
\cite{Giustina15,Shalm15,Scheidl10}, and the separation itself is nearly free
at telecom, $58$~m of fibre costing $0.012$~dB. The detection window is capped
by Eq.~(\ref{eq:timing}) rather than by the detector jitter: a nanosecond
window sits far inside the $193$~ns and far above the few picoseconds of
jitter, and the $100$~ps assumed for the dark-count estimate of
Sec.~\ref{sec:thresholds} is more comfortable still. A fixed window opened at
a fixed delay after the herald also closes the coincidence-time loophole
\cite{LarssonGill04,Larsson14}, under which a setting-dependent window admits
local models above the bound. What is not demonstrated on this platform is the
switching: Guerreiro \textit{et al.}\ did not switch settings at all.

Two effects we do not model can be bounded. A dark count on the herald
detector produces a trial with vacuum in both arms, indistinguishable from
upstream loss, so it multiplies $p$. A $1$~ns herald window at a $9.5$~MHz
pump admits only a hundredth of the dark rate. At the $2$~kHz herald rate of
Ref.~\cite{Guerreiro16}, a detector producing $100$ dark counts per second
then gives a false-herald fraction of $5\times10^{-4}$, costing $\Delta
S=-0.0016$; the $500$ per second measured in Ref.~\cite{Chang21} gives
$2.4\times10^{-3}$, costing $-0.008$, an eighth of the violation. The herald
window is a design parameter, not a detail. A misassigned photon number
likewise flips the parity outcome, so a probability $\epsilon$ of odd-order
misassignment rescales every correlator by $(1-2\epsilon)^2$, exactly as dark
counts do in Eq.~(\ref{eq:dark}), and the parity violation dies at
$\epsilon^*=\tfrac12[1-\sqrt{2/\max|S|_{\mu=1}}]=0.027$. Here $\epsilon$
excludes the dominant misassignment $1\to0$, which is detector inefficiency
and already carried by $\eta$; it stands for crosstalk and pulse-height
misclassification, a channel with no analogue in on--off detection. Several
omissions are second order and sign-definite, because the settings sit at a
maximum of $|S|$: local-oscillator intensity noise, displacement
miscalibration, a modulator setting error, an imperfect splitting ratio. All
can only reduce $|S|$. The last is quantified by the unbalanced-state scan of
Sec.~\ref{sec:practical}.

\subsection{Three practical points}
\label{sec:practical}

\emph{The wavelength is not the difficulty.} The $810$~nm entry of
Table~\ref{tab:hardware} uses a nanowire measured at $850$~nm and designed for
$940$, where nanowires reach $0.945$; the loophole-free tests at $810$~nm used
transition-edge sensors and \emph{measured} $\mu=0.786$ \cite{Giustina15}.
Both wavelengths are close enough that the choice should be made on the
source: telecom for the distance, the fibre and the coupling record
\cite{Liu22}; $810$~nm for detectors and sources that already exist
\cite{Giustina15,Pereira13}. What neither offers yet is the overlap of
Fig.~\ref{fig:frontier}.

\emph{Do not use the parity observable.} This is about the observable, not the
hardware: a number-resolving detector read out in on--off mode reaches
$\mu_c=0.8258$ and has no parity-flip channel, so number resolution costs
nothing; what costs is dichotomising by parity. The old objection to the
hardware --- that it costs efficiency --- no longer holds either, a segmented
nanowire now reaching $0.98$ at $1555$~nm while resolving $32$ photons
\cite{PNRsnspd25}. Against the observable, three arguments survive: its
threshold is twelve percentage points higher, its dark-count ceiling five
times lower (Sec.~\ref{sec:thresholds}), and a misassigned photon number
\emph{flips} its outcome, a first-order channel with no on--off analogue that
kills the violation at the $\epsilon^*$ of Sec.~\ref{sec:outside}. On the
idler the recommendation reverses: number resolution there vetoes a two-pair
herald outright, attacking $f_2$ at its root. We do not optimise over
dichotomic functions of $n$ more generally, and the family $x^{\hat n}$ leaves
room to.

\emph{Spend the effort on heralding, and on the interferometer.} Everything
else is close to its ceiling. A deterministic source does not yet help: the
best quantum-dot device delivers $0.712$ into single-mode fibre \cite{Ding25},
packaged versions far less \cite{Margaria25}, and quantum dots match poorly to
a coherent local oscillator, with wavepacket overlaps near $0.76$
\cite{Lam25}.

We also checked whether unbalancing the state helps, as it does for
polarisation entanglement. Replacing Eq.~(\ref{eq:psi1}) by
$\cos\theta|1,0\rangle-\sin\theta|0,1\rangle$ and optimising all four complex
settings, the threshold rises monotonically as $\theta$ leaves $\pi/4$:
$\mu_c=0.8258$ at $\cos^2\theta=\tfrac12$, then $0.8290$, $0.8329$, $0.8445$,
$0.8583$ and $0.8809$ at $\cos^2\theta=0.5627$, $0.5937$, $0.6545$, $0.7129$
and $0.7939$. Exchanging the modes maps $\theta$ to $\pi/2-\theta$, and the
rise is monotonic over the range we sampled, so the balanced state is optimal
there; we do not claim more. The rise is quadratic,
$\Delta\mu_c\simeq0.8\,\delta^2$ with $\delta=\cos^2\theta-\tfrac12$, which
doubles as a tolerance on the beam splitter that makes the state: a splitter
that is $52\!:\!48$ rather than balanced costs $3\times10^{-4}$ of threshold.
This is not in conflict with the protocols that do reach the Eberhard limit
\cite{Eberhard93,BC12,Alwehaibi25}: those unbalance vacuum against
$|1,1\rangle$, a superposition in which loss attenuates only one amplitude, so
the state drifts \emph{towards} the optimal Eberhard weight as loss increases.
Here loss attenuates both amplitudes equally and dumps the difference into an
incoherent vacuum term. The two are different resources, and only the former
buys anything.

\section{\texorpdfstring{Larger $N$}{Larger N}}
\label{sec:largeN}

The $N=1$ case is the one an experiment would attempt and the one the debate
of Sec.~\ref{sec:debate} is about, but Eq.~(\ref{eq:Pigen}) holds for every
$N$ and the two schemes differ there as well.

For $N\ge2$ the parity scheme does not violate CHSH even at unit efficiency
\cite{Wildfeuer07}, and loss makes it worse: the undisplaced settings give
$\Pi_x=(1-2\mu)^N$ for every pair, so the trivial value falls from $2$ as
$2|1-2\mu|^N$. The on--off scheme does violate, for every $N$ we tested. With
$+1$ assigned to no click,
\begin{equation}
  \max|S|=2.6884,\;2.0731,\;2.0108,\;2.0014,\;2.0001
  \label{eq:SN}
\end{equation}
for $N=1,\dots,5$ at unit efficiency. The violations for $N\ge2$ are small but
not artefacts: for $N\le4$ we re-evaluated $S$ at the optimal settings by
direct summation in a truncated Fock space, with no analytic expression, and
the two agree to machine precision. The optimal configuration is simple ---
for $N\ge2$ one setting on each side is the undisplaced measurement,
$\alpha=\beta=0$, and the other is a displacement of equal magnitude at the
two stations, $|\alpha'|=|\beta'|=0.635$ at $N=2$ and $0.694$ at $N=3$, with
relative phase $\arg(\alpha'^*\beta')=180^\circ/N$. Only $(\alpha'^*\beta')^N$
enters Eq.~(\ref{eq:Pigen}), so that phase is fixed only modulo $360^\circ/N$.
The phase itself, though, is not optional. Over the family with the two
displacements real and equal, $|S|=2-4e^{-R^2}R^{2N}/N!$ at
$\alpha'=\beta'=R$. That is strictly below $2$ for every finite $R>0$, and
reaches $2$ only in the two trivial limits: no displacement, and a
displacement large enough to make the detector click every time. The
antiparallel choice that is optimal at $N=1$ survives only at odd $N$.

Two qualifications are needed, and together they take most of the significance
out of the result.

First, by Eq.~(\ref{eq:CHident}) these are the Clauser--Horne violations of
Ref.~\cite{Wildfeuer07} in a different inequality; what is new is only the
dependence on efficiency, which that paper did not treat.

Second, these violations live at essentially unit efficiency. The $N=2$
requirement is $\mu_c=0.9230$, already above the $N=1$ on--off threshold by
nearly ten percentage points, and the $N=3$ requirement $\mu_c=0.9554$ exceeds
even the $N=1$ parity threshold that Sec.~\ref{sec:exp} argues is already out
of reach; $N=4$ and $N=5$ need more still. Equation~(\ref{eq:SN}) is therefore
a statement about the structure of the correlator, not about a feasible
experiment, and we do not claim more than the $N\le5$ we tested.

The direction is nonetheless consistent: the measurement that extracts less
information in principle is the one that survives realistic detection.

\section{Conclusion}
\label{sec:concl}

Whether a single photon delocalised over two modes can violate a Bell
inequality is not an open question of principle. The state is entangled, the
superselection objection was conditional on the absence of a phase reference,
and a local oscillator supplies one, provided the two oscillators are drawn
from a common source. A violation then certifies the nonlocality of the state
together with that reference. What is open is whether the violation can be
measured, a question about the apparatus.

We have reduced it to one number. The loss channel is displacement covariant,
so loss ahead of the displacement and loss at the detector are interchangeable
for any $N$. An experiment with symmetric arms therefore has a single overall
efficiency $\mu=p\eta$, and the requirement is $\mu>0.8258$ for on--off
detection or $\mu>0.9427$ for parity detection. Dark counts enter five to six
orders of magnitude below where they matter, the parity scheme dying at
$\nu^*=0.028$ per detector per window and the on--off scheme at $0.14$; phase
noise is comfortable by two and a half orders, and common-mode drift cancels
exactly. Mode matching is the one that is not free, and it costs $1.6$
percentage points of on--off threshold at the reported overlap.

The scheme to build is the simpler one, and by a wider margin than the ideal
analysis alone would show. On--off detection tolerates twelve percentage
points more loss and five times the dark counts, and it alone violates beyond
$N=1$ for the $N\le5$ we tested. The parity \emph{observable} buys nothing
here and costs a great deal, for a reason Eq.~(\ref{eq:lossparity}) makes
plain: a parity detector spends the usefulness of parity as its efficiency
falls, and has none left at $\eta=1/2$. The hardware is not the problem: a
number-resolving detector read out in on--off mode reaches the lower
threshold.

What stands between the analysis and an experiment is no longer efficiency.
The best demonstrated telecom coupling, with a detector two points from
perfect and an interferometer one can actually build, gives $\mu=0.862$,
comfortably past $\mu_c=0.8419$. The obstacle has moved, and
Eq.~(\ref{eq:xieff}) says where. The overlap that sets the threshold is not a
property of the local oscillator: it carries the weight of the heralded photon
in a single spectral mode, and that weight is bought by filtering, which is
paid for in the same heralding efficiency that supplies $\mu$. The two source
numbers in Table~\ref{tab:hardware} are the two ends of that trade, measured
in different laboratories. At $\mu=0.862$ the requirement is an effective
overlap of $0.92$, a single-mode weight of $0.85$ against a perfect local
oscillator. A source that reports both numbers, on one device, is what the
experiment is waiting for --- and after thirty years of argument, it is not
waiting on anything conceptual.

\section*{Disclosure of AI use}

Claude Opus 5 (Anthropic) was used for literature search, symbolic and
numerical verification of the results reported here, and for drafting and
editing the manuscript. All analytical results were independently checked
against brute-force Fock-space computation, and the author is responsible for
the content.


\begin{thebibliography}{99}

\bibitem{Sanders89} B.~C. Sanders, Phys. Rev. A \textbf{40}, 2417 (1989).
  \href{https://doi.org/10.1103/PhysRevA.40.2417}{doi:10.1103/PhysRevA.40.2417}
\bibitem{Boto00} A.~N. Boto, P. Kok, D.~S. Abrams, S.~L. Braunstein,
  C.~P. Williams, and J.~P. Dowling, Phys. Rev. Lett. \textbf{85}, 2733 (2000).
  \href{https://doi.org/10.1103/PhysRevLett.85.2733}{doi:10.1103/PhysRevLett.85.2733}
\bibitem{Lee02} H. Lee, P. Kok, and J.~P. Dowling,
  J. Mod. Opt. \textbf{49}, 2325 (2002).
  \href{https://doi.org/10.1080/0950034021000011536}{doi:10.1080/0950034021000011536}
\bibitem{Kapale05} K.~T. Kapale, L.~D. Didomenico, H. Lee, P. Kok, and
  J.~P. Dowling, Concepts of Physics \textbf{II}, 225 (2005).
  \href{https://doi.org/10.48550/arXiv.quant-ph/0507150}{doi:10.48550/arXiv.quant-ph/0507150}

\bibitem{CHSH69} J.~F. Clauser, M.~A. Horne, A. Shimony, and R.~A. Holt,
  Phys. Rev. Lett. \textbf{23}, 880 (1969).
  \href{https://doi.org/10.1103/PhysRevLett.23.880}{doi:10.1103/PhysRevLett.23.880}
\bibitem{CH74} J.~F. Clauser and M.~A. Horne,
  Phys. Rev. D \textbf{10}, 526 (1974).
  \href{https://doi.org/10.1103/PhysRevD.10.526}{doi:10.1103/PhysRevD.10.526}
\bibitem{Werner89} R.~F. Werner, Phys. Rev. A \textbf{40}, 4277 (1989).
  \href{https://doi.org/10.1103/PhysRevA.40.4277}{doi:10.1103/PhysRevA.40.4277}
\bibitem{Barrett02} J. Barrett, Phys. Rev. A \textbf{65}, 042302 (2002).
  \href{https://doi.org/10.1103/PhysRevA.65.042302}{doi:10.1103/PhysRevA.65.042302}
\bibitem{Gisin91} N. Gisin, Phys. Lett. A \textbf{154}, 201 (1991).
  \href{https://doi.org/10.1016/0375-9601(91)90805-I}{doi:10.1016/0375-9601(91)90805-I}
\bibitem{GisinPeres92} N. Gisin and A. Peres,
  Phys. Lett. A \textbf{162}, 15 (1992).
  \href{https://doi.org/10.1016/0375-9601(92)90949-M}{doi:10.1016/0375-9601(92)90949-M}
\bibitem{PopescuRohrlich92} S. Popescu and D. Rohrlich,
  Phys. Lett. A \textbf{166}, 293 (1992).
  \href{https://doi.org/10.1016/0375-9601(92)90711-T}{doi:10.1016/0375-9601(92)90711-T}
\bibitem{WJD07} H.~M. Wiseman, S.~J. Jones, and A.~C. Doherty,
  Phys. Rev. Lett. \textbf{98}, 140402 (2007).
  \href{https://doi.org/10.1103/PhysRevLett.98.140402}{doi:10.1103/PhysRevLett.98.140402}
\bibitem{Renou17} M.~O. Renou, D. Rosset, A. Martin, and N. Gisin,
  J. Phys. A \textbf{50}, 255301 (2017).
  \href{https://doi.org/10.1088/1751-8121/aa6f78}{doi:10.1088/1751-8121/aa6f78}
\bibitem{TWC91} S.~M. Tan, D.~F. Walls, and M.~J. Collett,
  Phys. Rev. Lett. \textbf{66}, 252 (1991).
  \href{https://doi.org/10.1103/PhysRevLett.66.252}{doi:10.1103/PhysRevLett.66.252}
\bibitem{Santos92} E. Santos, Phys. Rev. Lett. \textbf{68}, 894 (1992);
  S.~M. Tan, D.~F. Walls, and M.~J. Collett, \textit{ibid.} \textbf{68},
  895 (1992).
  \href{https://doi.org/10.1103/PhysRevLett.68.894}{doi:10.1103/PhysRevLett.68.894}
\bibitem{Hardy94} L. Hardy, Phys. Rev. Lett. \textbf{73}, 2279 (1994).
  \href{https://doi.org/10.1103/PhysRevLett.73.2279}{doi:10.1103/PhysRevLett.73.2279}
\bibitem{Peres95} A. Peres, Phys. Rev. Lett. \textbf{74}, 4571 (1995);
  \textit{ibid.} \textbf{76}, 2205(E) (1996).
  \href{https://doi.org/10.1103/PhysRevLett.74.4571}{doi:10.1103/PhysRevLett.74.4571}
\bibitem{Vaidman95} L. Vaidman, Phys. Rev. Lett. \textbf{75}, 2063 (1995).
  \href{https://doi.org/10.1103/PhysRevLett.75.2063}{doi:10.1103/PhysRevLett.75.2063}
\bibitem{GHZ95} D.~M. Greenberger, M.~A. Horne, and A. Zeilinger,
  Phys. Rev. Lett. \textbf{75}, 2064 (1995); L. Hardy, \textit{ibid.}
  \textbf{75}, 2065 (1995).
  \href{https://doi.org/10.1103/PhysRevLett.75.2064}{doi:10.1103/PhysRevLett.75.2064}
\bibitem{vanEnk05} S.~J. van Enk, Phys. Rev. A \textbf{72}, 064306 (2005).
  \href{https://doi.org/10.1103/PhysRevA.72.064306}{doi:10.1103/PhysRevA.72.064306}
\bibitem{Drezet06} A. Drezet, Phys. Rev. A \textbf{74}, 026301 (2006).
  \href{https://doi.org/10.1103/PhysRevA.74.026301}{doi:10.1103/PhysRevA.74.026301}
\bibitem{vanEnk06} S.~J. van Enk, Phys. Rev. A \textbf{74}, 026302 (2006).
  \href{https://doi.org/10.1103/PhysRevA.74.026302}{doi:10.1103/PhysRevA.74.026302}
\bibitem{Das21} T. Das, M. Karczewski, A. Mandarino, M. Markiewicz,
  B. Wo{\l}oncewicz, and M. {\.Z}ukowski, New J. Phys. \textbf{23},
  073042 (2021).
  \href{https://doi.org/10.1088/1367-2630/ac0ffe}{doi:10.1088/1367-2630/ac0ffe}
\bibitem{DasNoGo} T. Das, M. Karczewski, A. Mandarino, M. Markiewicz,
  B. Wo{\l}oncewicz, and M. {\.Z}ukowski, arXiv:2102.03254.
  \href{https://doi.org/10.48550/arXiv.2102.03254}{doi:10.48550/arXiv.2102.03254}
\bibitem{CD08} J.~J. Cooper and J.~A. Dunningham,
  New J. Phys. \textbf{10}, 113024 (2008).
  \href{https://doi.org/10.1088/1367-2630/10/11/113024}{doi:10.1088/1367-2630/10/11/113024}
\bibitem{DasComment} T. Das, M. Karczewski, A. Mandarino, M. Markiewicz,
  B. Wo{\l}oncewicz, and M. {\.Z}ukowski,
  New J. Phys. \textbf{24}, 038001 (2022).
  \href{https://doi.org/10.1088/1367-2630/ac55b1}{doi:10.1088/1367-2630/ac55b1}
\bibitem{Schlichtholz26} K. Schlichtholz, B. Wo{\l}oncewicz, T. Das,
  M. Markiewicz, and M. {\.Z}ukowski, Quantum \textbf{10}, 2084 (2026).
  \href{https://doi.org/10.22331/q-2026-04-28-2084}{doi:10.22331/q-2026-04-28-2084}

\bibitem{AS67} Y. Aharonov and L. Susskind, Phys. Rev. \textbf{155},
  1428 (1967).
  \href{https://doi.org/10.1103/PhysRev.155.1428}{doi:10.1103/PhysRev.155.1428}
\bibitem{BDSW06} S.~D. Bartlett, A.~C. Doherty, R.~W. Spekkens, and
  H.~M. Wiseman, Phys. Rev. A \textbf{73}, 022311 (2006).
  \href{https://doi.org/10.1103/PhysRevA.73.022311}{doi:10.1103/PhysRevA.73.022311}
\bibitem{BRS07} S.~D. Bartlett, T. Rudolph, and R.~W. Spekkens,
  Rev. Mod. Phys. \textbf{79}, 555 (2007).
  \href{https://doi.org/10.1103/RevModPhys.79.555}{doi:10.1103/RevModPhys.79.555}
\bibitem{DV07} J.~A. Dunningham and V. Vedral,
  Phys. Rev. Lett. \textbf{99}, 180404 (2007).
  \href{https://doi.org/10.1103/PhysRevLett.99.180404}{doi:10.1103/PhysRevLett.99.180404}
\bibitem{Ashhab09} S. Ashhab, K. Maruyama, {\v C}. Brukner, and F. Nori,
  Phys. Rev. A \textbf{80}, 062106 (2009).
  \href{https://doi.org/10.1103/PhysRevA.80.062106}{doi:10.1103/PhysRevA.80.062106}
\bibitem{BCB13} J.~B. Brask, R. Chaves, and N. Brunner,
  Phys. Rev. A \textbf{88}, 012111 (2013).
  \href{https://doi.org/10.1103/PhysRevA.88.012111}{doi:10.1103/PhysRevA.88.012111}
\bibitem{Abiuso22} P. Abiuso, T. Kriv{\'a}chy, E.-C. Boghiu, M.-O. Renou,
  A. Pozas-Kerstjens, and A. Ac{\'{\i}}n,
  Phys. Rev. Research \textbf{4}, L012041 (2022).
  \href{https://doi.org/10.1103/PhysRevResearch.4.L012041}{doi:10.1103/PhysRevResearch.4.L012041}

\bibitem{Paris96} M.~G.~A. Paris, Phys. Lett. A \textbf{217}, 78 (1996).
  \href{https://doi.org/10.1016/0375-9601(96)00339-8}{doi:10.1016/0375-9601(96)00339-8}
\bibitem{CG69} K.~E. Cahill and R.~J. Glauber,
  Phys. Rev. \textbf{177}, 1882 (1969).
  \href{https://doi.org/10.1103/PhysRev.177.1882}{doi:10.1103/PhysRev.177.1882}
\bibitem{BW96} K. Banaszek and K. W\'odkiewicz,
  Phys. Rev. Lett. \textbf{76}, 4344 (1996).
  \href{https://doi.org/10.1103/PhysRevLett.76.4344}{doi:10.1103/PhysRevLett.76.4344}
\bibitem{WV96} S. Wallentowitz and W. Vogel,
  Phys. Rev. A \textbf{53}, 4528 (1996).
  \href{https://doi.org/10.1103/PhysRevA.53.4528}{doi:10.1103/PhysRevA.53.4528}
\bibitem{BW99} K. Banaszek and K. W\'odkiewicz,
  Phys. Rev. Lett. \textbf{82}, 2009 (1999).
  \href{https://doi.org/10.1103/PhysRevLett.82.2009}{doi:10.1103/PhysRevLett.82.2009}
\bibitem{BRWK99} K. Banaszek, C. Radzewicz, K. W\'odkiewicz, and
  J.~S. Krasi\'nski, Phys. Rev. A \textbf{60}, 674 (1999).
  \href{https://doi.org/10.1103/PhysRevA.60.674}{doi:10.1103/PhysRevA.60.674}
\bibitem{Royer77} A. Royer, Phys. Rev. A \textbf{15}, 449 (1977).
  \href{https://doi.org/10.1103/PhysRevA.15.449}{doi:10.1103/PhysRevA.15.449}
\bibitem{MoyaCessa93} H. Moya-Cessa and P.~L. Knight,
  Phys. Rev. A \textbf{48}, 2479 (1993).
  \href{https://doi.org/10.1103/PhysRevA.48.2479}{doi:10.1103/PhysRevA.48.2479}
\bibitem{Wildfeuer07} C.~F. Wildfeuer, A.~P. Lund, and J.~P. Dowling,
  Phys. Rev. A \textbf{76}, 052101 (2007).
  \href{https://doi.org/10.1103/PhysRevA.76.052101}{doi:10.1103/PhysRevA.76.052101}
\bibitem{LJJ09} S.-W. Lee, H. Jeong, and D. Jaksch,
  Phys. Rev. A \textbf{80}, 022104 (2009).
  \href{https://doi.org/10.1103/PhysRevA.80.022104}{doi:10.1103/PhysRevA.80.022104}
\bibitem{Lee17} S.-Y. Lee, J. Park, J. Kim, and C. Noh,
  Phys. Rev. A \textbf{95}, 012134 (2017).
  \href{https://doi.org/10.1103/PhysRevA.95.012134}{doi:10.1103/PhysRevA.95.012134}
\bibitem{CB11} R. Chaves and J.~B. Brask,
  Phys. Rev. A \textbf{84}, 062110 (2011).
  \href{https://doi.org/10.1103/PhysRevA.84.062110}{doi:10.1103/PhysRevA.84.062110}
\bibitem{GargMermin87} A. Garg and N.~D. Mermin,
  Phys. Rev. D \textbf{35}, 3831 (1987).
  \href{https://doi.org/10.1103/PhysRevD.35.3831}{doi:10.1103/PhysRevD.35.3831}
\bibitem{Eberhard93} P.~H. Eberhard,
  Phys. Rev. A \textbf{47}, R747 (1993).
  \href{https://doi.org/10.1103/PhysRevA.47.R747}{doi:10.1103/PhysRevA.47.R747}
\bibitem{Bennet12} A.~J. Bennet, D.~A. Evans, D.~J. Saunders, C. Branciard,
  E.~G. Cavalcanti, H.~M. Wiseman, and G.~J. Pryde,
  Phys. Rev. X \textbf{2}, 031003 (2012).
  \href{https://doi.org/10.1103/PhysRevX.2.031003}{doi:10.1103/PhysRevX.2.031003}
\bibitem{Paterek11} T. Paterek, P. Kurzy\'nski, D.~K.~L. Oi, and
  D. Kaszlikowski, New J. Phys. \textbf{13}, 043027 (2011).
  \href{https://doi.org/10.1088/1367-2630/13/4/043027}{doi:10.1088/1367-2630/13/4/043027}
\bibitem{WV03} H.~M. Wiseman and J.~A. Vaccaro,
  Phys. Rev. Lett. \textbf{91}, 097902 (2003).
  \href{https://doi.org/10.1103/PhysRevLett.91.097902}{doi:10.1103/PhysRevLett.91.097902}
\bibitem{RudolphSanders01} T. Rudolph and B.~C. Sanders,
  Phys. Rev. Lett. \textbf{87}, 077903 (2001).
  \href{https://doi.org/10.1103/PhysRevLett.87.077903}{doi:10.1103/PhysRevLett.87.077903}
\bibitem{vanEnkFuchs02} S.~J. van Enk and C.~A. Fuchs,
  Phys. Rev. Lett. \textbf{88}, 027902 (2002).
  \href{https://doi.org/10.1103/PhysRevLett.88.027902}{doi:10.1103/PhysRevLett.88.027902}
\bibitem{BRST09} S.~D. Bartlett, T. Rudolph, R.~W. Spekkens, and
  P.~S. Turner, New J. Phys. \textbf{11}, 063013 (2009).
  \href{https://doi.org/10.1088/1367-2630/11/6/063013}{doi:10.1088/1367-2630/11/6/063013}
\bibitem{BCHKM02} J. Barrett, D. Collins, L. Hardy, A. Kent, and S. Popescu,
  Phys. Rev. A \textbf{66}, 042111 (2002).
  \href{https://doi.org/10.1103/PhysRevA.66.042111}{doi:10.1103/PhysRevA.66.042111}
\bibitem{Gill03} R.~D. Gill, in \textit{Foundations of Probability and
  Physics --- 2} (V\"axj\"o University Press, 2003), p.~179;
  arXiv:quant-ph/0301059.
  \href{https://doi.org/10.48550/arXiv.quant-ph/0301059}{doi:10.48550/arXiv.quant-ph/0301059}
\bibitem{Bierhorst15} P. Bierhorst,
  J. Phys. A \textbf{48}, 195302 (2015).
  \href{https://doi.org/10.1088/1751-8113/48/19/195302}{doi:10.1088/1751-8113/48/19/195302}
\bibitem{LarssonGill04} J.-\AA. Larsson and R.~D. Gill,
  Europhys. Lett. \textbf{67}, 707 (2004).
  \href{https://doi.org/10.1209/epl/i2004-10124-7}{doi:10.1209/epl/i2004-10124-7}
\bibitem{Larsson14} J.-\AA. Larsson,
  J. Phys. A \textbf{47}, 424003 (2014).
  \href{https://doi.org/10.1088/1751-8113/47/42/424003}{doi:10.1088/1751-8113/47/42/424003}
\bibitem{Hensen15} B. Hensen \textit{et al.},
  Nature \textbf{526}, 682 (2015).
  \href{https://doi.org/10.1038/nature15759}{doi:10.1038/nature15759}
\bibitem{Shalm15} L.~K. Shalm \textit{et al.},
  Phys. Rev. Lett. \textbf{115}, 250402 (2015).
  \href{https://doi.org/10.1103/PhysRevLett.115.250402}{doi:10.1103/PhysRevLett.115.250402}
\bibitem{Rosenfeld17} W. Rosenfeld, D. Burchardt, R. Garthoff, K. Redeker,
  N. Ortegel, M. Rau, and H. Weinfurter,
  Phys. Rev. Lett. \textbf{119}, 010402 (2017).
  \href{https://doi.org/10.1103/PhysRevLett.119.010402}{doi:10.1103/PhysRevLett.119.010402}
\bibitem{Scheidl10} T. Scheidl \textit{et al.},
  Proc. Natl. Acad. Sci. USA \textbf{107}, 19708 (2010).
  \href{https://doi.org/10.1073/pnas.1002780107}{doi:10.1073/pnas.1002780107}
\bibitem{Vivoli15} V. Caprara Vivoli, P. Sekatski, J.-D. Bancal, C.~C.~W. Lim,
B.~G. Christensen, A. Martin, R.~T. Thew, H. Zbinden, N. Gisin, and N.
Sangouard, Phys. Rev. A \textbf{91}, 012107 (2015).
  \href{https://doi.org/10.1103/PhysRevA.91.012107}{doi:10.1103/PhysRevA.91.012107}
\bibitem{Kun25} D. Kun, T. Str\"omberg, B. Daki\'c, P. Walther, and
  L.~A. Rozema, Optica \textbf{13}, 745 (2026).
  \href{https://doi.org/10.1364/OPTICA.586172}{doi:10.1364/OPTICA.586172}
\bibitem{BC12} J.~B. Brask and R. Chaves,
  Phys. Rev. A \textbf{86}, 010103(R) (2012).
  \href{https://doi.org/10.1103/PhysRevA.86.010103}{doi:10.1103/PhysRevA.86.010103}
\bibitem{Alwehaibi25} Y.~K. Alwehaibi, E. Mer, G.~J. Machado, S. Yu,
  I.~A. Walmsley, and R.~B. Patel,
  Phys. Rev. Research \textbf{7}, 043198 (2025).
  \href{https://doi.org/10.1103/xw66-nqfs}{doi:10.1103/xw66-nqfs}
\bibitem{Bjerrum23} A.~J.~E. Bjerrum, J.~B. Brask, J.~S. Neergaard-Nielsen,
  and U.~L. Andersen, Phys. Rev. A \textbf{107}, 052611 (2023).
  \href{https://doi.org/10.1103/PhysRevA.107.052611}{doi:10.1103/PhysRevA.107.052611}
\bibitem{Mycroft23} M.~E. Mycroft, T. McDermott, A. Buraczewski, and
  M. Stobi\'nska, Phys. Rev. A \textbf{107}, 012607 (2023).
  \href{https://doi.org/10.1103/PhysRevA.107.012607}{doi:10.1103/PhysRevA.107.012607}

\bibitem{Hessmo04} B. Hessmo, P. Usachev, H. Heydari, and G. Bj\"ork,
  Phys. Rev. Lett. \textbf{92}, 180401 (2004).
  \href{https://doi.org/10.1103/PhysRevLett.92.180401}{doi:10.1103/PhysRevLett.92.180401}
\bibitem{Babichev04} S.~A. Babichev, J. Appel, and A.~I. Lvovsky,
  Phys. Rev. Lett. \textbf{92}, 193601 (2004).
  \href{https://doi.org/10.1103/PhysRevLett.92.193601}{doi:10.1103/PhysRevLett.92.193601}
\bibitem{DAngelo06} M. D'Angelo, A. Zavatta, V. Parigi, and M. Bellini,
  Phys. Rev. A \textbf{74}, 052114 (2006).
  \href{https://doi.org/10.1103/PhysRevA.74.052114}{doi:10.1103/PhysRevA.74.052114}
\bibitem{Fuwa15} M. Fuwa, S. Takeda, M. Zwierz, H.~M. Wiseman, and
  A. Furusawa, Nat. Commun. \textbf{6}, 6665 (2015).
  \href{https://doi.org/10.1038/ncomms7665}{doi:10.1038/ncomms7665}
\bibitem{Monteiro15} F. Monteiro, V. Caprara Vivoli, T. Guerreiro, A. Martin,
  J.-D. Bancal, H. Zbinden, R.~T. Thew, and N. Sangouard,
  Phys. Rev. Lett. \textbf{114}, 170504 (2015).
  \href{https://doi.org/10.1103/PhysRevLett.114.170504}{doi:10.1103/PhysRevLett.114.170504}
\bibitem{Liu22} W.-Z. Liu, Y.-Z. Zhang, Y.-Z. Zhen, M.-H. Li,
  Y. Liu, J. Fan, F. Xu, Q. Zhang, and J.-W. Pan,
  Phys. Rev. Lett. \textbf{129}, 050502 (2022).
  \href{https://doi.org/10.1103/PhysRevLett.129.050502}{doi:10.1103/PhysRevLett.129.050502}
\bibitem{MeyerScott17} E. Meyer-Scott, N. Montaut, J. Tiedau, L. Sansoni,
  H. Herrmann, T.~J. Bartley, and C. Silberhorn,
  Phys. Rev. A \textbf{95}, 061803(R) (2017).
  \href{https://doi.org/10.1103/PhysRevA.95.061803}{doi:10.1103/PhysRevA.95.061803}
\bibitem{Guerreiro16} T. Guerreiro \textit{et al.},
  Phys. Rev. Lett. \textbf{117}, 070404 (2016).
  \href{https://doi.org/10.1103/PhysRevLett.117.070404}{doi:10.1103/PhysRevLett.117.070404}
\bibitem{Caspar22} P. Caspar, E. Oudot, P. Sekatski, N. Maring, A. Martin,
  N. Sangouard, H. Zbinden, and R.~T. Thew, Quantum \textbf{6}, 671 (2022).
  \href{https://doi.org/10.22331/q-2022-03-22-671}{doi:10.22331/q-2022-03-22-671}
\bibitem{Giustina15} M. Giustina \textit{et al.},
  Phys. Rev. Lett. \textbf{115}, 250401 (2015).
  \href{https://doi.org/10.1103/PhysRevLett.115.250401}{doi:10.1103/PhysRevLett.115.250401}

\bibitem{Reddy20} D.~V. Reddy, R.~R. Nerem, S.~W. Nam, R.~P. Mirin, and
  V.~B. Verma, Optica \textbf{7}, 1649 (2020).
  \href{https://doi.org/10.1364/OPTICA.400751}{doi:10.1364/OPTICA.400751}
\bibitem{EsmaeilZadeh20} I. Esmaeil Zadeh \textit{et al.},
  ACS Photonics \textbf{7}, 1780 (2020).
  \href{https://doi.org/10.1021/acsphotonics.0c00433}{doi:10.1021/acsphotonics.0c00433}
\bibitem{Chang21} J. Chang \textit{et al.},
  APL Photonics \textbf{6}, 036114 (2021).
  \href{https://doi.org/10.1063/5.0039772}{doi:10.1063/5.0039772}
\bibitem{Korzh20} B. Korzh \textit{et al.},
  Nat. Photonics \textbf{14}, 250 (2020).
  \href{https://doi.org/10.1038/s41566-020-0589-x}{doi:10.1038/s41566-020-0589-x}
\bibitem{Fukuda11} D. Fukuda \textit{et al.},
  Opt. Express \textbf{19}, 870 (2011).
  \href{https://doi.org/10.1364/OE.19.000870}{doi:10.1364/OE.19.000870}
\bibitem{Los24} J.~W.~N. Los, M. Sidorova \textit{et al.},
  APL Photonics \textbf{9}, 066101 (2024).
  \href{https://doi.org/10.1063/5.0204340}{doi:10.1063/5.0204340}
\bibitem{Pereira13} M. da Cunha Pereira, F.~E. Becerra, B.~L. Glebov, J. Fan,
  S.~W. Nam, and A. Migdall, Opt. Lett. \textbf{38}, 1609 (2013).
  \href{https://doi.org/10.1364/OL.38.001609}{doi:10.1364/OL.38.001609}
\bibitem{Ding25} X. Ding \textit{et al.},
  Nat. Photonics \textbf{19}, 387 (2025).
  \href{https://doi.org/10.1038/s41566-025-01639-8}{doi:10.1038/s41566-025-01639-8}
\bibitem{Margaria25} N. Margaria \textit{et al.},
  Nat. Commun. \textbf{16}, 7553 (2025).
  \href{https://doi.org/10.1038/s41467-025-62712-y}{doi:10.1038/s41467-025-62712-y}
\bibitem{Lam25} H. Lam \textit{et al.},
  Quantum Sci. Technol. \textbf{10}, 045061 (2025).
  \href{https://doi.org/10.1088/2058-9565/ae0a7a}{doi:10.1088/2058-9565/ae0a7a}
\bibitem{Svarc23} V. {\v S}varc, M. Nov{\'a}kov{\'a}, M. Dudka, and
  M. Je{\v z}ek, Opt. Express \textbf{31}, 12562 (2023).
  \href{https://doi.org/10.1364/OE.480569}{doi:10.1364/OE.480569}
\bibitem{PNRsnspd25} C. Ding, X. Zhang, J. Xiong, Y. Xiao, T. Zhang,
  J. Huang, H. Xu, X. Liu, L. You, Z. Wang, and H. Li,
  ACS Photonics \textbf{12}, 4924 (2025).
  \href{https://doi.org/10.1021/acsphotonics.5c00508}{doi:10.1021/acsphotonics.5c00508}

\end{thebibliography}
\end{document}